\documentclass[rmp,twocolumn]{revtex4}

\usepackage{graphicx}
\usepackage{bm}
\usepackage{url}

\begin{document}
%%%%%%%%%%%%%%%%%%%%%%%%%%%%%%%%%%%%%%%%%%%%%%%%%%%%%%%%%%%%%%%%%%%%%

\title{Applications of statistical mechanics to economics:  Entropic origin of the probability distributions of money, income, and energy consumption}

\author{Victor M.~Yakovenko}

\affiliation{Department of Physics, University of
  Maryland, College Park, Maryland 20742-4111, USA}

\date{29 April 2012 post on arXiv}
%\date{v.4, 28 January 2012}

\begin{abstract}
This Chapter is written for the Festschrift celebrating the 70th birthday of the distinguished economist Duncan Foley from the New School for Social Research in New York.  This Chapter reviews applications of statistical physics methods, such as the principle of entropy maximization, to the probability distributions of money, income, and global energy consumption per capita.  The exponential probability distribution of wages, predicted by the statistical equilibrium theory of a labor market developed by Foley in 1996, is supported by empirical data on income distribution in the USA for the majority (about 97\%) of population.  In addition, the upper tail of income distribution (about 3\% of population) follows a power law and expands dramatically during financial bubbles, which results in a significant increase of the overall income inequality.  A mathematical analysis of the empirical data clearly demonstrates the two-class structure of a society, as pointed out Karl Marx and recently highlighted by the Occupy Movement.  Empirical data for the energy consumption per capita around the world are close to an exponential distribution, which can be also explained by the entropy maximization principle. \\

\textsf{\normalsize ``Money, it's a gas.'' Pink Floyd, 
\textit{Dark Side of the Moon}}
\end{abstract}

\maketitle

%%%%%%%%%%%%%%%%%%%%%%%%%%%%%%%%%%%%%%%%%%%%%%%%%%%%%%%%%%%%%%%
\section{How I met Duncan Foley}
\label{Sec:Foley}
%%%%%%%%%%%%%%%%%%%%%%%%%%%%%%%%%%%%%%%%%%%%%%%%%%%%%%%%%%%%%%%

Although I am a theoretical physicist, I have been always interested in economics and, in particular, in applications of statistical physics to economics.  These ideas first occurred to me when I was an undergraduate student at the Moscow Physical-Technical Institute in Russia and studied statistical physics for the first time.  However, it was not until 2000 when I published my first paper on this subject \cite{Dragulescu-2000}, joining the emerging movement of econophysics \cite{Shea-2005,Farmer-2005,Hogan-2005}.  At that time, I started looking for economists who may be interested in a statistical approach to economics.  I attended a seminar by Eric Slud, a professor of mathematics at the University of Maryland, who independently explored similar ideas and eventually published them in \textcite{Silver-2002}.  In this paper, I saw a reference to the paper by \textcite{Foley-1994}.  So, I contacted Duncan in January 2001 and invited him to give a talk at the University of Maryland, which he did in March.  Since then, our paths have crossed many times.  I visited the New School for Social Research in New York several times, and we also met and discussed at the Santa Fe Institute, where I was spending a part of my sabbatical in January--February 2009, hosted by Doyne Farmer.  During these visits, I also met other innovative economists at the New School, professors Anwar Shaikh and Willi Semler, as well as Duncan's Ph.D.\ student at that time Mishael Milakovi\'c, who is now a professor of economics at the University of Bamberg in Germany.

As it is explained on the first page of \textcite{Foley-1994}, Duncan learned statistical physics by taking a course on statistical thermodynamics at Dartmouth College.  For me, the  papers by \textcite{Foley-1994,Foley-1996,Foley-1999} conjure an image of people who 
meet on a bridge trying to reach for the same ideas by approaching from the opposite banks, physics and economics.  Over time, I learned about other papers where economists utilize statistical and entropic ideas, e.g.\ \textcite{Golan-1994,Molico-2006,Aoki-2006}, but I also realized how exceedingly rare statistical approach is among mainstream economists.  Duncan is one of the very few innovative economists who has deeply studied statistical physics and attempted to make use of it in economics.

One of the puzzling social problems is persistent economic inequality among the population in any society.  In statistical physics, it is well known that identical (``equal'') molecules in a gas spontaneously develop a widely unequal distribution of energies as a result of random energy transfers in molecular collisions.  By analogy, very unequal probability distributions can spontaneously develop in an economic system as a result of random interactions between economic agents.  This is the main idea of the material presented in this Chapter.

First, I will briefly review the basics of statistical physics and then discuss and compare the applications of these ideas to economics in \textcite{Foley-1994,Foley-1996,Foley-1999} and in my papers, from the first paper \cite{Dragulescu-2000} to the most recent review papers \cite{Yakovenko-2009,Banerjee-2010} and books \cite{Cockshott-2009,Yakovenko-2011}.
As we shall see, statistical and entropic ideas have a multitude of applications to the probability distributions of money, income, and global energy consumption.  The latter topic has relevance to the ongoing debate on the economics of global warming \cite{Rezai-2012}.  Duncan Foley and Eric Smith of the Santa Fe Institute have also published a profound study of the connection between phenomenological thermodynamics in physics and the utility formalism in economics \cite{Smith-2008}.  This paper will not be reviewed here because of a limited volume of this Chapter.  Unfortunately, there is still no complete understanding of a connection between statistical mechanics and phenomenological thermodynamics as applied to economics.  These issues remain open for a future study.  I will only focus on statistical mechanics in this Chapter.

%%%%%%%%%%%%%%%%%%%%%%%%%%%%%%%%%%%%%%%%%%%%%%%%%%%%%%%%%%%%%%%
\section{Entropy and the Boltzmann-Gibbs distribution of energy in physics}
\label{Sec:BGphysics}
%%%%%%%%%%%%%%%%%%%%%%%%%%%%%%%%%%%%%%%%%%%%%%%%%%%%%%%%%%%%%%%

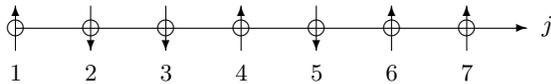
\begin{figure}

\setlength{\unitlength}{1cm}
\begin{picture}(8.5,1.5)(-0.7,-0.7)

\multiput(0,0)(1,0){7}{\circle{0.2}}
\put(0,-0.3){\vector(0,1){0.6}}
\put(1,0.3){\vector(0,-1){0.6}}
\put(2,0.3){\vector(0,-1){0.6}}
\put(3,-0.3){\vector(0,1){0.6}}
\put(4,0.3){\vector(0,-1){0.6}}
\put(5,-0.3){\vector(0,1){0.6}}
\put(6,-0.3){\vector(0,1){0.6}}

\put(-0.075,-0.7){1}
\put(0.925,-0.7){2}
\put(1.925,-0.7){3}
\put(2.925,-0.7){4}
\put(3.925,-0.7){5}
\put(4.925,-0.7){6}
\put(5.925,-0.7){7}

\put(0,0){\vector(1,0){6.8}}
\put(7,-0.05){$j$}

\end{picture}

\caption{A quantum paramagnet, where each atom has the spin $s=1/2$ and is in one of the two possible states ($q=2$): $k=1=\uparrow$ or $k=2=\downarrow$.  The atoms are labeled consecutively by the integer index $j=1,2,\ldots,N$.}
\label{fig:lattice}
\end{figure}

In this section, I briefly review the basics of statistical physics, starting from a discrete model of a quantum paramagnet.  Let us consider a collection of $N$ atoms and label the individual atoms with the integer index $j=1,2,\ldots,N$.  It is convenient to visualize the atoms as sitting on a lattice, as shown in Fig.~\ref{fig:lattice}.  Each atom has an internal quantum degree of freedom called the spin $s$ and can be in one of the $q$ discrete states.\footnote{In quantum mechanics, $s$ takes integer or half-integer values, and $q=2s+1$, but this is not important here.}  These states are labeled by the index $k=1,2,\ldots,q$.  The system is placed in a external magnetic field, so each discrete state has a different energy $\varepsilon_k$.  
Then, the state of an atom $j$ can be characterized by the energy $\varepsilon_j$ that this atom has. (Throughout the paper, I use the indices $i$ and $j$ to label individual atoms or agents, and the index $k$ to label possible states of the atoms or agents.)  A simple case with $q=2$ is illustrated in Fig.~\ref{fig:lattice}, where each atom can be in one of the two possible states $k=1=\uparrow$ or $k=2=\downarrow$, which are depicted as the spin-up and spin-down states.  Alternatively, one can interpret each atom as a bit of computer memory with two possible states 0 and 1.  For $q=8$, each atoms would correspond to a byte with 8 possible values.  The atoms can be also interpreted as economic agents, where the index $k$ indicates an internal state of each agent.

%%%%%%%%%%%%%%%%%%%%%%%%%%%%%%%%%%%%%%%%%%%%%%%%%%%%%%%%%%%%%%%
%%%%%%%%%%%%%%%%%%%%%%%%%%%%%%%%%%%%%%%%%%%%%%%%%%%%%%%%%%%%%%%

For a given configuration of atoms, let us count how many atoms are in each state $k$ and denote these numbers as $N_k$, so that $\sum_{k=1}^qN_k=N$.  Let us introduce the multiplicity $\Omega$, which is the number of different atomic configurations that have the same set of the numbers $N_k$.  Calculating $\Omega$, we treat the atoms as distinguishable, because they are localized on distinguishable lattice sites labeled by the index $j$ in Fig.~\ref{fig:lattice}.  For example, let us consider the case of $N=2$ and $q=2$, i.e.\ two atoms with two possible states.  The set $N_1=1$ and $N_2=1$ can be realized in two possible ways, so $\Omega=2$.  In one configuration, the first atom $j=1$ is in the state $k=1=\uparrow$ with the energy $\varepsilon_1=\varepsilon_\uparrow$, whereas the second atom $j=2$ is in the state $k=2=\downarrow$ with the energy $\varepsilon_2=\varepsilon_\downarrow$.  In another configuration, the states of the atoms are reversed.  On the other hand, the set $N_1=2$ and $N_2=0$ can be realized in only one way, where both atoms are in the state $k=1=\uparrow$ with the energies $\varepsilon_1=\varepsilon_2=\varepsilon_\uparrow$, so $\Omega=1$.  In general, the multiplicity $\Omega$ can be calculated combinatorially as the number of different placements of $N$ atoms into $q$ boxes with the fixed occupations $N_k$ of each box: 
\begin{equation}
  \Omega=\frac{N!}{N_1!\,N_2!\,N_3!\,\ldots\,N_q!}.
\label{multiplicity}
\end{equation}
The logarithm of multiplicity is called the entropy $S=\ln\Omega$. Using the Stirling approximation for the factorials in the limit of large numbers $N_k$, we obtain
\begin{equation}
  S=N\ln N - \sum_{k=1}^q N_k\ln N_k =
  -\sum_{k=1}^q N_k\ln\left(\frac{N_k}{N}\right).
\label{S}
\end{equation}

In the absence of further information, we assume that all microscopic configurations of atoms are equally probable.  Then, the probability of observing a certain set of the numbers $N_k$ is proportional to the number of possible microscopic realizations of this set, i.e.\ to the multiplicity $\Omega$.  So,  the most probable set of the numbers $N_k$ is the one that maximizes $\Omega$ or $S$ subject to certain constraints.  The typical constraints for a closed system are that the total number of atoms $N$ and the total energy $E$ are fixed:
\begin{equation}
  N=\sum_{k=1}^q N_k, \qquad E=\sum_{k=1}^q \varepsilon_k N_k
  = \sum_{j=1}^N \varepsilon_j.
\label{constraints}
\end{equation}
To implement the constraints, we introduce the Lagrange multipliers $\alpha$ and $\beta$ and construct the modified entropy
\begin{equation}
  \tilde S = S + \alpha\sum_{k=1}^q N_k - \beta\sum_{k=1}^q \varepsilon_k N_k.
\label{S'}
\end{equation}
Maximization of $S$ is achieved by setting the derivatives $\partial\tilde S/\partial N_k$ to zero for each $N_k$.  Substituting Eq.~(\ref{S}) into Eq.~(\ref{S'}) and taking the derivatives,\footnote{Notice that $N=\sum_k N_k$ in Eq.~(\ref{S}) should be also differentiated with respect to $N_k$.} we find that the occupation numbers $P(\varepsilon_k)=N_k/N$ for the states $k$ depend exponentially on the energies of these states $\varepsilon_k$
\begin{equation}
  P(\varepsilon_k)=\frac{N_k}{N}=e^{\alpha-\beta \varepsilon_k}
  =e^{-(\varepsilon_k-\mu)/T}.
\label{P(e)}
\end{equation}
Here the parameters $T=1/\beta$ and $\mu=\alpha T$ are called the temperature and the chemical potential.  The values of $\alpha$ and $\beta$ are determined by substituting Eq.~(\ref{P(e)}) into Eq.~(\ref{constraints}) and satisfying the constraints for given $N$ and $E$.  The temperature $T$ is equal to the average energy per particle $T\sim\langle\varepsilon\rangle=E/N$, up to a numerical coefficient of the order of one.  The interpretation of Eq.~(\ref{P(e)}) is that the probability for an atom to occupy a state with the energy $\varepsilon$ depends exponentially on the energy: $P(\varepsilon)\propto\exp(-\varepsilon/T)$.  

This consideration can be generalized to the case where the label $k$ becomes a continuous variable.  In this case, the sums in Eqs.~(\ref{S}), (\ref{constraints}), and (\ref{S'}) are replaced by integrals with an appropriate measure of integration.  The typical physical example is a gas of atoms moving inside a large box.  In this case, the atoms have closely-spaced quantized energy levels, and the number of quantum states within a momentum interval $\Delta p$ and a space interval $\Delta x$ is $\Delta p\,\Delta x/h$, where $h$ is the Planck constant.  Then, the measure of integration in Eqs.~(\ref{S}), (\ref{constraints}), and (\ref{S'}) becomes the element of volume of the phase space $dp\,dx/h$ in one-dimensional case and $d^3p\,d^3r/h^3$ in three-dimensional case.  The energy in Eq.~(\ref{P(e)}) becomes the kinetic energy of an atom $\varepsilon=p^2/2m$, which is bounded from below $\varepsilon\geq0$.

Eq.~(\ref{P(e)}) is the fundamental law of equilibrium statistical physics \cite{Wannier-book}, known as the Boltzmann-Gibbs distribution.\footnote{In physics, the elementary particles are indistinguishable and obey the Fermi-Dirac or Bose-Einstein statistics, rather than the Boltzmann statistics.  In contrast, the economic agents are distinguishable because of their human identity.  Thus, the Boltzmann statistics for distinguishable objects is appropriate in this case.}  However, one can see that the above derivation is really an exercise in theory of probabilities and, as such, is not specific to physics.  Thus, it can be applied to statistical ensembles of different nature, subject to constraints similar to Eq.~(\ref{constraints}).  Interpreting the discrete states of the atoms as bits of information, one arrives to the Shannon entropy in the theory of information.  Entropy concepts have been applied to such disparate fields as ecology \cite{Banavar-2010,Blundell-2011} and neuroscience \cite{Wen-2009,Varshney-2006}.

The economy is also a big statistical system consisting of a large number of economic agents.  Thus, it is tempting to apply the formalism presented above to the economy as well.  However, there are different ways of implementing such an analogy.  In the next section, I briefly summarize an analogy developed in \textcite{Foley-1994,Foley-1996,Foley-1999}.

%%%%%%%%%%%%%%%%%%%%%%%%%%%%%%%%%%%%%%%%%%%%%%%%%%%%%%%%%%%%%%%
\section{Statistical equilibrium theory of markets by Duncan Foley}
\label{Sec:markets}
%%%%%%%%%%%%%%%%%%%%%%%%%%%%%%%%%%%%%%%%%%%%%%%%%%%%%%%%%%%%%%%

In \textcite{Foley-1994}, Duncan proposed a statistical equilibrium theory of markets.  This theory is an alternative to the conventional competitive equilibrium theory originated by Walras and by Marshall.  In the conventional theory, an auctioneer collects orders from buyers and sellers and determines an equilibrium price that clears the market, subject to budget constraints and utility preferences of the agents.  In this theory, all transactions take place at the same price, but it is not quite clear how the real markets would actually converge to this price.  In contrast, Duncan proposed a theory where is a probability distribution of trades at different prices, and market clearing is achieved only statistically.

\textcite{Foley-1994} studied an ensemble of $N$ economic agents who trade $n$ types of different commodities.  The state of each agent $j=1,2,3,\ldots,N$ is characterized by the $n$-component vector $\bm x_j=(x_j^{(1)},x_j^{(2)},x_j^{(3)},\ldots,x_j^{(n)})$.  Each component of this vector represents a possible trade that the agent $j$ is willing to perform with a given commodity.   Positive, negative, and zero values of $x_j$ represent an increase, a decrease, and no change in the stock of a given commodity for the agent $j$.  All trades in the system are subject to the global constraint
\begin{equation}
  \sum\nolimits_j \bm x_j=0,
\label{xj=0}
\end{equation}
which represents conservation of commodities in the process of trading, i.e.\ commodities are only transferred between the agents.  An increase $x_i>0$ of a commodity stock for an agent $i$ must be compensated by a decrease $x_j<0$ for another agent $j$, so that the algebraic sum (\ref{xj=0}) of all trades is zero.  However, Eq.~(\ref{xj=0}) does not require bilateral balance of transactions between pairs of agents and allows for multilateral trades.

The vector $\bm x_j$ can be considered as an $n$-dimensional generalization of the variable $\varepsilon_j$ introduced in Sec.~\ref{Sec:BGphysics} to characterize the state of each atom.  Different kinds of trades $\bm x_k$ are labeled by the index $k$, and the number of agents doing the trade $\bm x_k$ is denoted as $N_k$.  Then the constraint (\ref{xj=0}) can be rewritten as
\begin{equation}
  \sum\nolimits_k \bm x_k \, N_k =0.
\label{xk=0}
\end{equation}
For a given set of $N_k$, the multiplicity (\ref{multiplicity}) gives the number of different assignments of individual agents to the trades $\bm x_k$, such that the numbers $N_k$ are fixed.  Maximizing the entropy (\ref{S}) subject to the constraints (\ref{xk=0}), we arrive to an analog of Eq.\ (\ref{P(e)}), which now has the form
\begin{equation}
  P(\bm x_k)=\frac{N_k}{N}=c\,e^{-\bm\pi\cdot\bm x_k}.
\label{P(x)}
\end{equation}
Here $c$ is a normalization constant, and $\bm\pi$ is the $n$-component vector of Lagrange multipliers introduced to satisfy the constraints (\ref{xk=0}).
\textcite{Foley-1994} interpreted $\bm\pi$ as the vector of entropic prices.  The probability for an agent to perform the set of trades $\bm x$ depends exponentially on the volume of the trades: $P(\bm x)\propto\exp(-\bm\pi\cdot\bm x)$.

In \textcite{Foley-1996}, the general theory \cite{Foley-1994} was applied to a simple labor market.  In this model, there are two classes on agents: employers (firms) and employees (workers).  They trade in two commodities ($n=2$): $x^{(1)}=w$ is wage, supplied by the firms and taken by the workers,
and $x^{(2)}=l$ is labor, supplied by the workers and taken by the firms.   For each worker, the offer set includes the line $\bm x=(w>w_0,-1)$, where $-1$ is the fixed offer of labor in exchange for any wage $w$ greater than a minimum wage $w_0$.  The offer set also includes the point $\bm x=(0,0)$, which offers no labor and no wage, i.e.\ the state of unemployment.  For each firm, the offer set is the line $\bm x=(-K,l>l_0)$, where $-K$ is the fixed amount of capital spent on paying wages in exchange for the amount of labor $l$ greater than a minimum value $l_0$.

A mathematically similar model was also developed in \textcite{Foley-1999} as a result of conversations with Perry Mehrling.  This model also has two commodities ($n=2$), one of which is interpreted as money and another as a financial asset, such as a bond or a treasury bill.

Let us focus on the labor market model by \textcite{Foley-1996}.  According to Eq.~(\ref{P(x)}), the model predicts the exponential probability distributions for the wages $w$ received by the workers and for the labor $l$ employed by the firms.  Let us try to compare these predictions with empirical data for the real economy.  The probability distribution of wages can be compared with income distribution, for which a lot of data is available and which will be discussed in Sec.~\ref{Sec:income}.  On the other hand, the distribution of labor employed by the firms can be related to the distribution of firm sizes, as measured by their number of employees.  However, \textcite{Foley-1996} makes an artificial simplifying assumption that each firm spends the same amount of capital $K$ on labor.  This is a clearly unrealistic assumption because of the great variation in the amount of capital among the firms, including the capital spent on labor.  Thus, let us focus only on the probability distribution of wages, unconditional on the distribution of labor.  To obtain this distribution, we take a sum in Eq.~(\ref{P(x)}) over the values of $x^{(2)}$, while keeping a fixed value for $x^{(1)}$.  In physics jargon, we ``integrate out'' the degree of freedom $x^{(2)}=l$ and thus obtain the unconditional probability distribution of the remaining degree of freedom $x^{(1)}=w$, which is still exponential
\begin{equation}
  P(w)= c\, e^{-w/T_w}.
\label{P(w)}
\end{equation}
Here $c$ is a normalization constant, and $T_w=1/\pi^{(1)}$ is the wage temperature.

Now let us discuss how the constraint (\ref{xk=0}) is satisfied with respect to wages.  The model has $N_f$ firms, which supply the total capital for wages $W=KN_f$, which enters as a negative term into the sum (\ref{xk=0}).  Since only the total amount matters in the constraint, we can take $W$ as an input parameter of the model.  Given that the unemployed workers have zero wage $w=0$, the constraint (\ref{xk=0}) can be rewritten as 
\begin{equation}
  \sum\nolimits_{w_k>w_0} w_k\,N_k =W, 
\label{wk=W}
\end{equation}
where the sum is taken over the employed workers, whose total number we denote as $N_e$.  The average wage per employed worker is $\langle w\rangle=W/N_e$.  Using Eq.~(\ref{P(w)}) and replacing summation over $k$ by integration over $w$ in  Eq.~(\ref{wk=W}), we relate $\langle w\rangle$ and $T_w$
\begin{equation}
  \langle w\rangle=\frac{W}{N_e}=\frac{\int_{w_0}^\infty w\,P(w)\,dw}
  {\int_{w_0}^\infty P(w)\,dw}, 
  \quad T_w = \langle w\rangle-w_0.
\label{<w>}
\end{equation}
So, the wage temperature $T_w$ is a difference of the average wage per employed worker and the minimal wage.

The model by \textcite{Foley-1996} also has unemployed workers, whose number $N_u$ depends on  the measure of statistical weight assigned to the state with $w=0$.  Because this measure is an input parameter of the model, one might as well take $N_u$ as an input parameter.  I will not further discuss the  number of unemployed and will focus on the distribution of wages among the employed workers.  It will be shown in Sec.~\ref{Sec:income} that the exponential distribution of wages (\ref{P(w)}) indeed agrees with the actual empirical data on income distribution for the majority of population.

%%%%%%%%%%%%%%%%%%%%%%%%%%%%%%%%%%%%%%%%%%%%%%%%%%%%%%%%%%%%%%%
\section{Statistical mechanics of money}
\label{Sec:BGmoney}
%%%%%%%%%%%%%%%%%%%%%%%%%%%%%%%%%%%%%%%%%%%%%%%%%%%%%%%%%%%%%%%

The theory presented in Sec.~\ref{Sec:markets} focused on market transactions and identified the states of the agents with the possible vectors of transactions $\bm x_j$.  However, an analogy with the Boltzmann-Gibbs formalism can be also developed in a different way, as proposed by \textcite{Dragulescu-2000}.  In this paper, the states of the agents were identified with the amounts of a commodity they hold, i.e.\ with stocks of a commodity, rather than with fluxes of a commodity.  This analogy is closer to the original physical picture presented in Sec.~\ref{Sec:BGphysics}, where the states of the atoms were identified with the amounts of energy $\varepsilon_j$ held by the atoms (i.e.\ the stocks of energy), rather than with the amounts of energy transfer in collisions between the atoms (i.e.\ the fluxes of energy).  In this section, I summarize the theory developed by \textcite{Dragulescu-2000}.

Similarly to Sec.~\ref{Sec:markets}, let us consider an ensemble of $N$ economic agents and characterize their states by the amounts (stocks) of commodities they hold.  Let us introduce one special commodity called money,\footnote{To simplify consideration, I use only one monetary instrument, commonly known to everybody as ``money''.} whereas all other commodities are assumed to be physically consumable, such as food, consumer goods, and services.  Money is fundamentally different from consumable goods, because it is an artificially created, non-consumable object.  I will not dwell on historical origins and various physical implementations of money, but will proceed straight to the modern fiat money.  Fundamentally, money is bits of information (digital balances) assigned to each agent, so money represents an informational layer of the economy, as opposed to the physical layer consisting of various manufactured and consumable goods.  Although money is not physically consumable, and well-being of the economic agents is ultimately determined by the physical layer, nevertheless money plays an extremely important role in the modern economy.  Many economic crises were caused by monetary problems, not by physical problems.  In the economy, monetary subsystem interacts with physical subsystem, but the two layers cannot transform into each other because of their different nature.  For this reason, an increase in material production does not result in an automatic increase in money supply.

Let us denote money balances of the agents $j=1,2,\ldots,N$ by $m_j$.
Ordinary economic agents can only receive money from and give money to other agents, and are not permitted to ``manufacture'' money, e.g.\ to print dollar bills.  The agents can grow apples on trees, but cannot grow money on trees.  Let us consider an economic transaction between the agents $i$ and $j$.  When the agent $i$ pays money $\Delta m$ to the agent $j$ for some goods or services, the money balances of the agents change as
\begin{eqnarray}
  && m_i\;\rightarrow\; m_i'=m_i-\Delta m,
\nonumber \\
  && m_j\;\rightarrow\; m_j'=m_j+\Delta m.
\label{transfer}
\end{eqnarray}
The total amount of money of the two agents before and after the
transaction remains the same
\begin{equation}
  m_i+m_j=m_i'+m_j',
\label{conservation}
\end{equation}
i.e.\ there is a local conservation law for money.  The transfer of money (\ref{transfer}) is analogous to the transfer of energy in molecular collisions, and Eq.~(\ref{conservation}) is analogous to conservation of energy.  Conservative models of this kind are also studied in some economic literature \cite{Kiyotaki-1993,Molico-2006}.

Enforcement of the local conservation law (\ref{conservation}) is crucial for successful functioning of money.  If the agents were permitted to ``manufacture'' money, they would be printing money and buying all goods for nothing, which would be a disaster.  In a barter economy, one consumable good  is exchanged for another (e.g.\ apples for oranges), so physical contributions of both agents are obvious.  However, this becomes less obvious in a monetary economy, where goods are exchanged for money and money for goods.  The purpose of the conservation law (\ref{transfer}) is to ensure that an agent can buy goods from the society only if he or she contributed something useful to the society and received monetary payment for these contributions.  Money is an accounting device, and, indeed, all accounting rules are based on the conservation law (\ref{transfer}).  

Unlike ordinary economic agents, a central bank or a central government, who issued the fiat money in the first place, can inject money into the economy, thus changing the total amount of money in the system.  This process is analogous to an influx of energy into a system from external sources.  As long as the rate of money influx is slow compared with the relaxation rate in the economy, the system remains in a quasi-stationary statistical equilibrium with slowly changing parameters.  This situation is analogous to slow heating of a kettle, where the kettle has a well defined, but slowly increasing, temperature at any moment of time.  Following the long-standing tradition in the economic literature, this section studies the economy in equilibrium, even though the real-world economy may be totally out of whack.  An economic equilibrium implies a monetary equilibrium too, so, for these idealized purposes, we consider a model where the central authorities do not inject additional money, so the total amount of money $M$ held by all ordinary agents in the system is fixed.  In this situation, the local conservation law (\ref{conservation}) becomes the global conservation law for money
\begin{equation}
  \sum\nolimits_j m_j=M.
\label{mj=0}
\end{equation}
It is important, however, that we study the system in a statistical equilibrium, as opposed to a mechanistic equilibrium envisioned by Walras and Marshall.  For this reason, the global constraint (\ref{mj=0}) is actually not as crucial for the final results, as it might seem.  In physics, we usually start with the idealization of a closed system, where the total energy $E$ in Eq.~(\ref{constraints}) is fixed.  Then, after deriving the Boltzmann-Gibbs distribution (\ref{P(e)}) and the concept of temperature, we generalize the consideration to an open system in contact with a thermal reservoir at a given temperature.  Even though the total energy of the open system is not conserved any more, the atoms still have the same Boltzmann-Gibbs distribution (\ref{P(e)}).  Similarly, routine daily operations of the central bank (as opposed to outstanding interventions, such as the recent quantitative easing by the Federal Reserve Bank) do not necessarily spoil the equilibrium distribution of money obtained for a closed system. 

Another potential problem with conservation of money is debt, which will be discussed in more detail in Sec.~\ref{Sec:debt}.  To simplify initial consideration, we do not allow agents to have debt in this section.  Thus, by construction, money balances of the agents cannot drop below zero. i.e.\ $m_i\geq0$ for all $i$.  Transaction (\ref{transfer}) takes place only when an agent has enough money to pay the price, i.e.\ $m_i\geq\Delta m$.  An agent with $m_i=0$ cannot buy goods from other agents, but can receive money for delivering goods or services to them.  Most econophysics models, see reviews by \textcite{Yakovenko-2009,QuantFinance-1,QuantFinance-2}, and some economic models \cite{Kiyotaki-1993,Molico-2006} do not consider debt, so it is not an uncommon simplifying assumption.  In this approach, the agents are liquidity-constrained.  As the recent economic crisis brutally reminded, liquidity constraint rules the economy, even though some players tend to forget about it during debt binges in the bubbles.

The transfer of money in Eq.~(\ref{transfer}) from one agent to another represents payment for delivered goods and services, i.e.\ there is an implied counterflow of goods for each monetary transaction.  However, keeping track explicitly of the stocks and fluxes of consumable goods would be a very complicated task.  One reason is that many goods, e.g.\ food and other supplies, and most services, e.g.\ getting a haircut or going to a movie, are not tangible and disappear after consumption.  Because they are not conserved, and also because they are measured in many different physical units, it is not practical to keep track of them.  In contrast, money is measured in the same unit (within a given country with a single currency) and is conserved in local transactions (\ref{conservation}), so it is straightforward to keep track of money.  

Thus, I choose to keep track only of the money balances of the agents $m_j$, but not of the other commodities in the system.  This situation can be visualized as follows.  Suppose money accounts of all agents are kept on a central computer server.  A computer operator observes the multitude of money transfers (\ref{transfer}) between the accounts, but he or she does not have information about the reasons for these transfers and does not know what consumable goods were transferred in exchange.    Although purposeful and rational for individual agents, these transactions look effectively random for the operator.  (Similarly, in statistical physics, each atom follows deterministic equations of motion, but, nevertheless, the whole system is effectively random.)  Out of curiosity, the computer operator records a snapshot of the money balances $m_j$ of all agents at a given time.  Then the operator counts the numbers of agents $N_k$ who have money balances in the intervals between $m_k$ and $m_k+m_*$, where $m_*$ is a reasonably small money window, and the index $k$ labels the money interval.  The operator wonders what is the general statistical principle that governs the relative occupation numbers $P(m_k)=N_k/N$.  In the absence of additional information, it is reasonable to derive these numbers by maximizing the multiplicity $\Omega$ and the entropy $S$ of the money distribution in Eqs.~(\ref{multiplicity}) and (\ref{S}), subject to the constraint $\sum_k m_k N_k=M$ in Eq.~(\ref{mj=0}).  The result is given by the Boltzmann-Gibbs distribution for money 
\begin{equation}
  P(m)=c\,e^{-m/T_m},
\label{P(m)}
\end{equation}
where $c$ is a normalizing constant, and $T_m$ is the money
temperature.  Similarly to Eq.~(\ref{<w>}), the temperature is obtained from the constraint (\ref{mj=0}) and is equal to the average amount of money per agent: $T_m=\langle m\rangle=M/N$.

%%%%%%%%%%%%%%%%%%%%%%%%%%%%%%%%%%%%%%%%%%%%%%%%%%%%%%%%%%%%%%%
\begin{figure}
\includegraphics[width=0.9\linewidth]{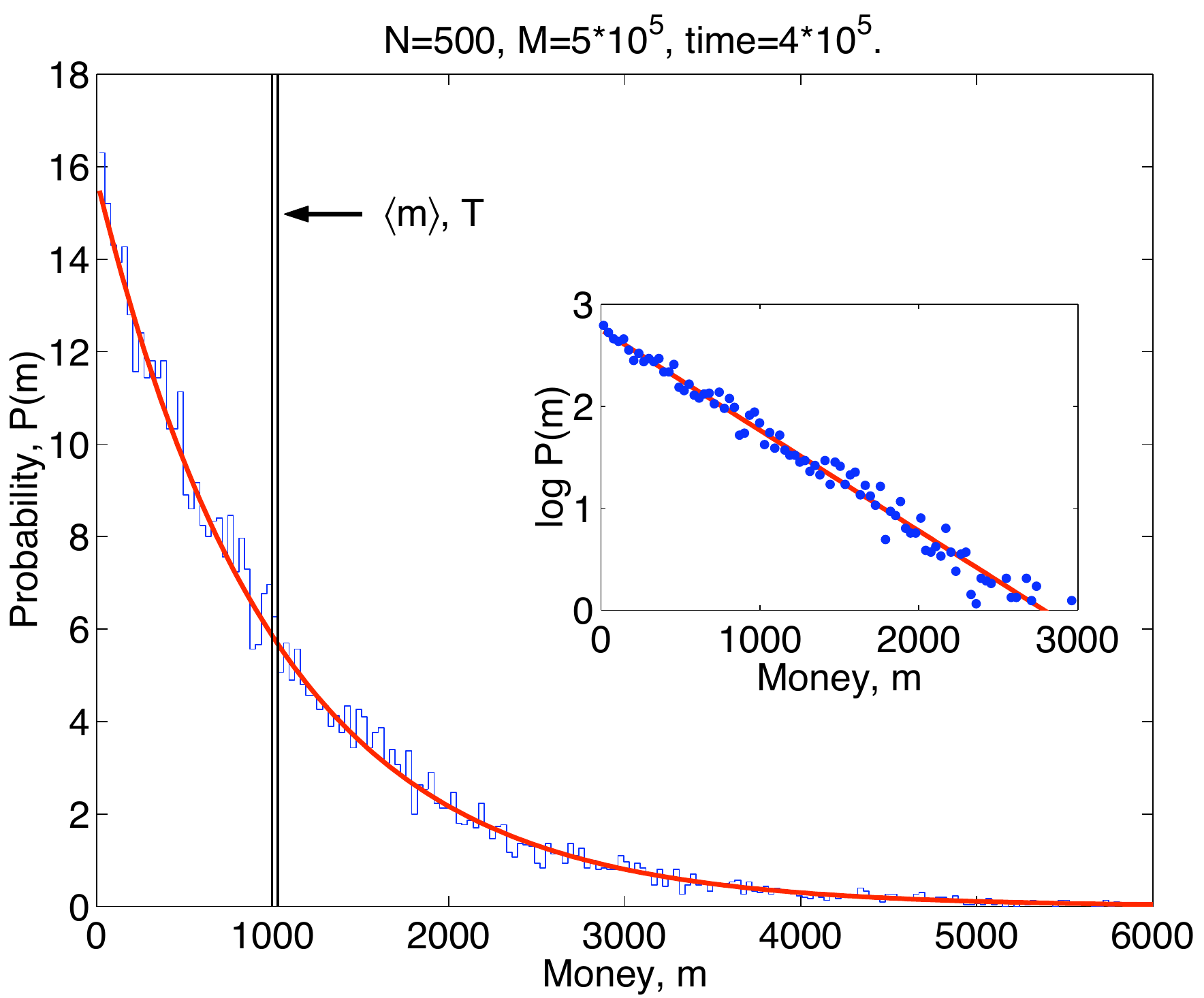}
\caption{\textit{Histogram and points:} The stationary probability
  distribution of money $P(m)$ obtained in computer simulations of the 
  random transfer models
  \cite{Dragulescu-2000}.  \textit{Solid curves:} Fits to the exponential
  distribution (\ref{P(m)}).  \textit{Vertical line:} The initial 
  distribution of money.}
\label{Fig:money}
\end{figure}
%%%%%%%%%%%%%%%%%%%%%%%%%%%%%%%%%%%%%%%%%%%%%%%%%%%%%%%%%%%%%%%

The statistical approach not only predicts the equilibrium distribution of money (\ref{P(m)}), but also allows us to study how the probability distribution of money $P(m,t)$ evolves in time $t$ toward the equilibrium.  \textcite{Dragulescu-2000} performed computer simulations of money transfers between the agents in different models.  Initially all agents were given the same amount of money, say, \$1000.  Then, a pair of agents $(i,j)$ was randomly selected, the amount $\Delta m$ was transferred from one agent to another, and the process
was repeated many times.  Time evolution of the probability
distribution of money $P(m,t)$ is shown in computer animation videos by \textcite{Chen-2007,Wright-2007}.  After a transitory period, money distribution converges to the stationary form shown in Fig.\ \ref{Fig:money}.  As expected, the distribution is well fitted by the exponential function (\ref{P(m)}).

In the simplest model considered by \textcite{Dragulescu-2000}, the transferred amount $\Delta m=\$1$ was constant.  Computer animation \cite{Chen-2007} shows that the initial distribution of money first broadens to a symmetric Gaussian curve, typical for a diffusion process.  Then, the distribution starts
to pile up around the $m=0$ state, which acts as the impenetrable
boundary, because of the imposed condition $m\geq0$.  As a result,
$P(m)$ becomes skewed (asymmetric) and eventually reaches the
stationary exponential shape, as shown in Fig.~\ref{Fig:money}.  The
boundary at $m=0$ is analogous to the ground-state energy in physics, i.e.\ the lowest possible energy of a physical system.  Without this boundary condition, the probability distribution of money would not reach a stationary shape.  Computer animations \cite{Chen-2007,Wright-2007} also show how the entropy (\ref{S}) of the money distribution, defined as $S/N=-\sum_k P(m_k)\ln P(m_k)$, grows from the initial value $S=0$, where all agents have the same money, to
the maximal value at the statistical equilibrium.

\textcite{Dragulescu-2000} performed simulations of several models with different rules for money transfers $\Delta m$.  As long as the rules satisfy the time-reversal symmetry \cite{Dragulescu-2000}, the stationary distribution is always the exponential one (\ref{P(m)}), irrespective of initial conditions and details of transfer rules.  However, this symmetry is violated by the multiplicative rules of transfer, such as the proportional rule $\Delta m=\gamma m_i$ \cite{Ispolatov-1998,Angle-1986}, the saving propensity \cite{Chakraborti-2000}, and the negotiable prices \cite{Molico-2006}.  These models produce Gamma-like distributions, as well as a power-law tail for a random distribution of saving propensities \cite{Chatterjee-2007}.  Despite some mathematical differences, all these models demonstrate spontaneous development of a highly unequal probability distribution of money as a result of random money transfers between the agents \cite{QuantFinance-1,QuantFinance-2}.  The money transfer models can be also formulated in the language of commodity trading between the agents optimizing their utility functions \cite{A.S.Chakrabarti-2009}, which is closer to the traditional language of the economic literature.  More involved agent-based simulations were developed by \textcite{Wright-2005,Wright-2009} and demonstrated emergence of a two-classes society from the initially equal agents.  This work was further developed in the book by \textcite{Cockshott-2009}, integrating economics, computer science, and physics.  Empirical data on income distribution discussed in Sec.~\ref{Sec:income} show direct evidence for the two-class society.

Let us compare the results of Secs.\ \ref{Sec:markets} and \ref{Sec:BGmoney}.  In the former, the Boltzmann-Gibbs formalism is applied to the fluxes of commodities, and the distribution of wages (\ref{P(w)}) is predicted to be exponential.  In the latter, the Boltzmann-Gibbs formalism is applied to the stocks of commodities, and the distribution of money balances (\ref{P(m)}) is predicted to be exponential.  There is no contradiction between these two approaches, and both fluxes and stocks may have exponential distributions simultaneously, although this is not necessarily required.  These papers were motivated by different questions.  The papers by \textcite{Foley-1994,Foley-1996,Foley-1999} were motivated by the long-standing question of how markets determine prices between different commodities in a decentralized manner.  The paper by \textcite{Dragulescu-2000} was motivated by another long-standing question of how inequality spontaneously develops among equal agents with equal initial endowments of money.  In the neoclassical thinking, equal agents with equal initial endowments should forever stay equal, which contradicts everyday experience.  Statistical approach argues that the state of equality is fundamentally unstable, because it has a very low entropy.  The law of probabilities \cite{Farjoun-1983} leads to the exponential distribution, which is highly unequal, but stable, because it maximizes the entropy.  Some papers in the economic literature modeled the distributions of both prices and money balances within a common statistical framework \cite{Molico-2006}.

It would be very interesting to compare the theoretical prediction (\ref{P(m)}) with empirical data on money distribution.  Unfortunately, it is very difficult to obtain such data.  The distribution of balances on deposit accounts in a big enough bank would be a reasonable approximation for the distribution of money among the population. However, such data are not publicly available.  In contrast, plenty of data on income distribution are available from the tax agencies and surveys of population.  These data will be compared with the theoretical prediction (\ref{P(w)}) in Sec.~\ref{Sec:income}.  However, before that, I will discuss in the next section how the results of this section would change when the agents are permitted to have debt.

%%%%%%%%%%%%%%%%%%%%%%%%%%%%%%%%%%%%%%%%%%%%%%%%%%%%%%%%%%%%%%%
\section{Models of debt}
\label{Sec:debt}
%%%%%%%%%%%%%%%%%%%%%%%%%%%%%%%%%%%%%%%%%%%%%%%%%%%%%%%%%%%%%%%

From the standpoint of individual economic agents, debt may be considered as negative money.  When an agent borrows money from a bank,\footnote{Here we treat the bank as being outside of the system consisting of ordinary agents, because we are interested in money distribution among these agents.  The debt of the agents is an asset for the bank, and deposits of cash into the bank are liabilities of the bank \cite{McConnell-book}.} the cash balance of the agent (positive money) increases, but the agent also acquires an equal debt obligation (negative money), so the total balance (net worth) of the agent
remains the same.  Thus, the act of money borrowing still satisfies a
generalized conservation law of the total money (net worth), which is
now defined as the algebraic sum of positive (cash $M$) and negative
(debt $D$) contributions: $M-D=M_b$, where $M_b$ is the original amount of money in the system, the monetary base \cite{McConnell-book}.  When an agent needs to buy a product at a price $\Delta m$ exceeding his money balance $m_i$, the agent is now permitted to borrow the difference from a bank.  After the transaction, the new balance of the agent becomes negative: $m_i'=m_i-\Delta m<0$.  The local conservation law (\ref{transfer}) and (\ref{conservation}) is still satisfied, but now it involves negative values of $m$.  Thus, the consequence of debt is not a violation of the conservation law, but a modification of the boundary condition by permitting negative balances $m_i<0$, so $m=0$ is not the ground state any more.

If the computer simulation with $\Delta m=\$1$ described in Sec.~\ref{Sec:BGmoney} is repeated without any restrictions on the debt of the agents, the probability distribution of money $P(m,t)$ never stabilizes, and the system never reaches a stationary state.  As time goes on, $P(m,t)$ keeps spreading in a Gaussian manner unlimitedly toward $m=+\infty$ and $m=-\infty$.  Because of the generalized conservation law, the first moment of the algebraically defined money $m$ remains constant $\langle m\rangle=M_b/N$.  It means that some agents become richer with positive balances $m>0$ at the expense of other agents going further into debt with negative balances $m<0$.

Common sense, as well as the current financial crisis, indicate that an economic system cannot be stable if unlimited debt is permitted.  In this case, the agents can buy goods without producing anything in exchange by simply going into unlimited debt.  Eq.~(\ref{P(m)}) is not applicable in this case for the following mathematical reason.  The normalizing coefficient in Eq.~(\ref{P(m)}) is $c=1/Z$, where $Z$ is the partition function
\begin{equation}
  Z=\sum\nolimits_k e^{-m_k/T_m},
\label{Z}
\end{equation}
and the sum is taken over all permitted states $k$ of the agents.  When debt is not permitted, and money balances are limited to non-negative values $m_k\geq0$, the sum in Eq.~(\ref{Z}) converges, if the temperature is selected to be positive $T_m>0$.  This is the case in physics, where kinetic energy takes non-negative values $\varepsilon\geq0$, and there is a ground state with the lowest possible energy, i.e.\ energy is bounded from below.\footnote{When the energy spectrum $\varepsilon_k$ is bounded both from above and below, the physical temperature $T$ may take either positive or negative values.  For a quantum paramagnet discussed in Sec~\ref{Sec:BGphysics}, the negative temperature $T$ corresponds to the inverse population, where the higher energy levels $\varepsilon_k$ have higher population $N_k$ than the lower energy levels.}  In contrast, when debt is permitted and is not limited by any constraints, the sum in the partition function (\ref{Z}) over both positive and negative values of $m_k$ diverges for any sign of the temperature $T_m$.

%%%%%%%%%%%%%%%%%%%%%%%%%%%%%%%%%%%%%%%%%%%%%%%%%%%%%%%%%%%%%%%
\begin{figure}
\includegraphics[width=0.9\linewidth]{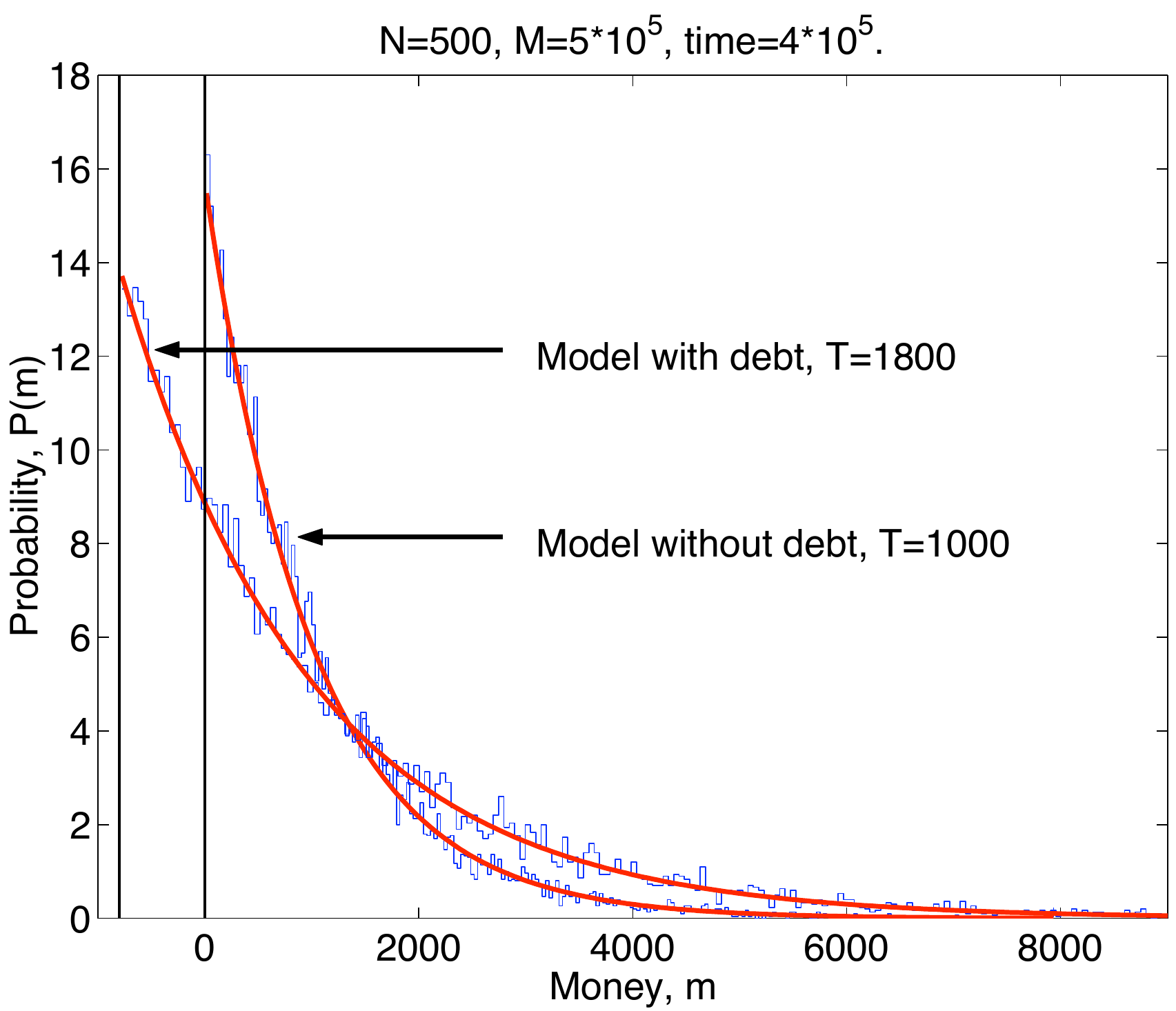}
\caption{\textit{Histograms:} Stationary distributions of money with
  and without debt \cite{Dragulescu-2000}.  The debt is limited to $m_d=800$.
  \textit{Solid curves:} Fits to the exponential distributions with the money
  temperatures $T_m=1800$ and $T_m=1000$.}
\label{Fig:debt}
\end{figure}
%%%%%%%%%%%%%%%%%%%%%%%%%%%%%%%%%%%%%%%%%%%%%%%%%%%%%%%%%%%%%%%

It is clear that some sort of a modified boundary condition has to be imposed in order to prevent unlimited growth of debt and to ensure overall stability and equilibrium in the system.  \textcite{Dragulescu-2000} considered a simple model where the maximal debt of each agent is limited to $m_d$.  In this model, $P(m)$ again has the exponential shape, but with the new boundary condition at $m=-m_d$ and the higher money temperature $T_d=m_d+M_b/N$, as shown in Fig.~\ref{Fig:debt}.  This result is analogous to Eq.~(\ref{<w>}) with the substitution $w_0\to-m_d$.  By allowing the agents to go into debt up to $m_d$, we effectively increase the amount of money available to each agent by $m_d$.

\begin{figure}
\includegraphics[width=\linewidth]{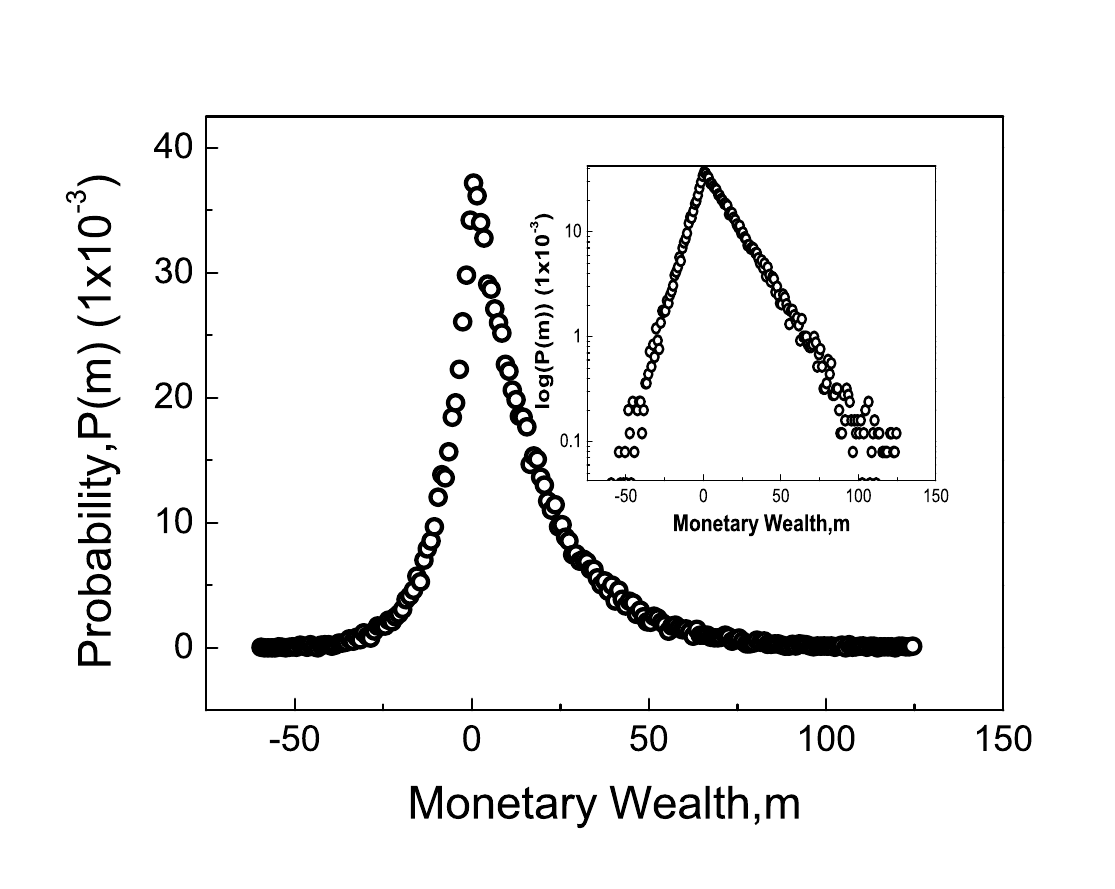}
\caption{The stationary distribution of money for the required reserve ratio 
  $R=0.8$ \cite{Xi-Ding-Wang-2005}.  The distribution is exponential for both 
  positive and negative money with different temperatures $T_+$ and $T_-$,
  as shown in the inset on log-linear scale.}
\label{Fig:reserve}
\end{figure}

\textcite{Xi-Ding-Wang-2005} considered a more realistic boundary condition, where a constraint is imposed on the total debt of all agents in the system, rather than on individual debt of each agent.  This is accomplished via the required reserve ratio $R$ \cite{McConnell-book}.  Banks are required by law to set aside a fraction $R$ of the money deposited into bank accounts, whereas the remaining fraction $1-R$ can be lent to the agents.  If the initial amount of money in the system (the money base) is $M_b$, then, with repeated lending and borrowing, the total amount of positive money available to the agents increases up to $M=M_b/R$, where the factor $1/R$ is called the money multiplier \cite{McConnell-book}.  This is how ``banks create money''.  This extra money comes from the increase in the total debt $D=M_b/R-M_b$ of the agents.  Now we have two constraints: on the positive money $M$ and on the maximal debt $D$.  Applying the principle of maximal entropy, we find two exponential distributions, for positive and negative money, characterized by two different temperatures, $T_+=M_b/RN$ and $T_-=M_b(1-R)/RN$.  This result was confirmed in computer simulations by \textcite{Xi-Ding-Wang-2005}, as shown in Fig.\ \ref{Fig:reserve}.

%%%%%%%%%%%%%%%%%%%%%%%%%%%%%%%%%%%%%%%%%%%%%%%%%%%%%%%%%%%%%%%
%%%%%%%%%%%%%%%%%%%%%%%%%%%%%%%%%%%%%%%%%%%%%%%%%%%%%%%%%%%%%%%

However, in the real economy, the reserve requirement is not effective in
stabilizing debt in the system, because it applies only to retail banks insured by FDIC, but not to investment banks.  Moreover, there are alternative instruments of debt, including derivatives, credit default swaps, and various other unregulated ``financial innovations''.  As a result, the total debt is not limited in practice and can reach catastrophic proportions.  \textcite{Dragulescu-2000} studied a model with different interest rates for deposits into and loans from a bank, but without an explicit limit on the total debt.  Computer simulations by \textcite{Dragulescu-2000} show that, depending on the choice of parameters, the total amount of money in circulation either increases or decreases in time, but never stabilizes.  The interest amplifies the destabilizing effect of debt, because positive balances become even more positive and negative even more negative.  Arguably, the current financial crisis was caused by the enormous debt accumulation, triggered by subprime mortgages and financial derivatives based on them.  The lack of restrictions was justified by a misguided notion that markets will always arrive to an equilibrium, as long as government does not interfere with them.  However, an equilibrium is possible only when the proper boundary conditions are imposed.  Debt does not stabilize by itself without enforcement of boundary conditions.

Bankruptcy is a mechanism for debt stabilization.  Interest rates are meaningless without a mechanism for triggering bankruptcy.  Bankruptcy erases the debt of an agent (the negative money) and resets the balance to zero.  However, somebody else (a bank or a lender) counted this debt as a positive asset, which also becomes erased.  In the language of physics, creation of debt is analogous to particle-antiparticle generation (creation of positive and negative money), whereas cancellation of debt corresponds to particle-antiparticle annihilation (annihilation of positive and negative money).  The former and the latter dominate during booms and busts and correspond to monetary expansion and contraction.

Besides fiat money created by central governments, numerous attempts have been made to create alternative community money from scratch \cite{Hokkaido}.  In such a system, when an agent provides goods or services to another agent, their accounts are credited with positive and negative tokens, as in Eq.\ (\ref{transfer}).  Because the initial money base $M_b=0$ is zero in this case, the probability distribution of money $P(m)$ is symmetric with respect to positive and negative $m$, i.e.\ ``all money is credit''.  Unless boundary conditions are imposed, the money distribution $P(m,t)$ would never stabilize in this system.  Some agents would try to accumulate unlimited negative balances by consuming goods and services and not contributing anything in return, thus undermining the system.

%%%%%%%%%%%%%%%%%%%%%%%%%%%%%%%%%%%%%%%%%%%%%%%%%%%%%%%%%%%%%%%
\section{Empirical data on income distribution}
\label{Sec:income}
%%%%%%%%%%%%%%%%%%%%%%%%%%%%%%%%%%%%%%%%%%%%%%%%%%%%%%%%%%%%%%%

%%%%%%%%%%%%%%%%%%%%%%%%%%%%%%%%%%%%%%%%%%%%%%%%%%%%%%%%%%%%%%%
\begin{figure}
\includegraphics[width=0.9\linewidth]{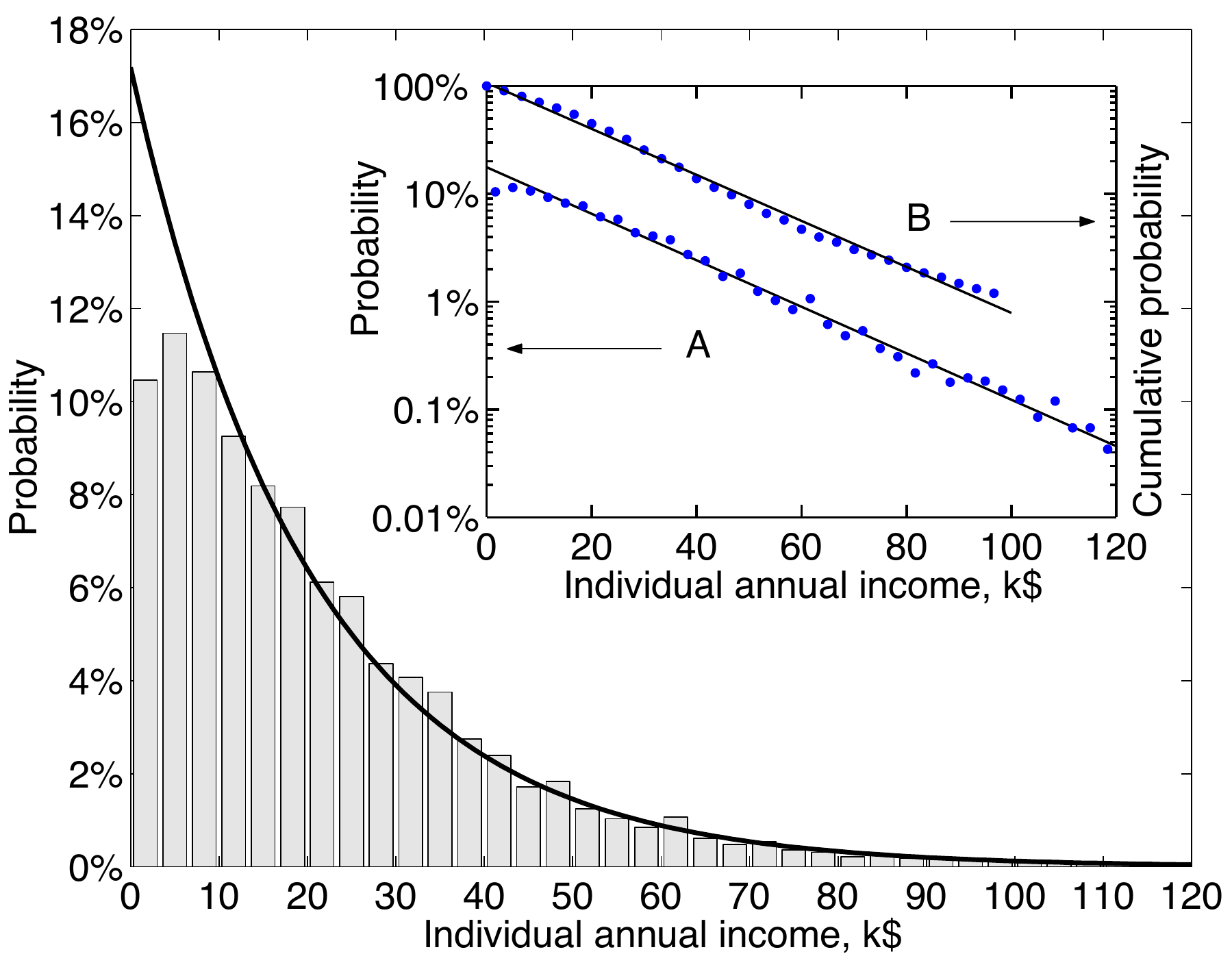}
\caption{  
  \textit{Histogram:} Probability distribution of individual income from the 
  US Census Bureau data for 1996 \cite{Dragulescu-2001a}.  
  \textit{Solid line:} Fit to the exponential law.  
  \textit{Inset plot A:} The same with the logarithmic vertical scale.  
  \textit{Inset plot B:} Cumulative probability distribution
  of individual income from the PSID data for 1992.}
\label{fig:census}
\end{figure}
%%%%%%%%%%%%%%%%%%%%%%%%%%%%%%%%%%%%%%%%%%%%%%%%%%%%%%%%%%%%%%%

In this section, the theoretical prediction (\ref{P(w)}) for the probability distribution of wages will be compared with the empirical data on income distribution in the USA.  The data from the US Census Bureau and the Internal Revenue Service (IRS) were analyzed by \textcite{Dragulescu-2001a}.  One plot from this paper is shown in Fig.~\ref{fig:census}.  The data agree very well with the exponential distribution, and the wage temperature $T_w=20.3$ k\$/year for 1996 is obtained from the fit.  Eq.~(\ref{P(w)}) also has the parameter $w_0$, which represents the minimal wage.  Although the data deviate from the exponential distribution in Fig.~\ref{fig:census} for incomes below 5 k\$/year, the probability density at $w=0$ is still non-zero, $P(0)\neq0$.  One would expect difficulties in collecting reliable data for very low incomes.  Given limited accuracy of the data in this range and the fact that the deviation occurs well below the average income $T_w$, we can set $w_0=0$ for the practical purposes of fitting the data.  Although there is a legal requirement for minimal hourly wage, there is no such requirement for annual wage, which depends on the number of hours worked.

\begin{figure}
\includegraphics[width=0.9\linewidth]{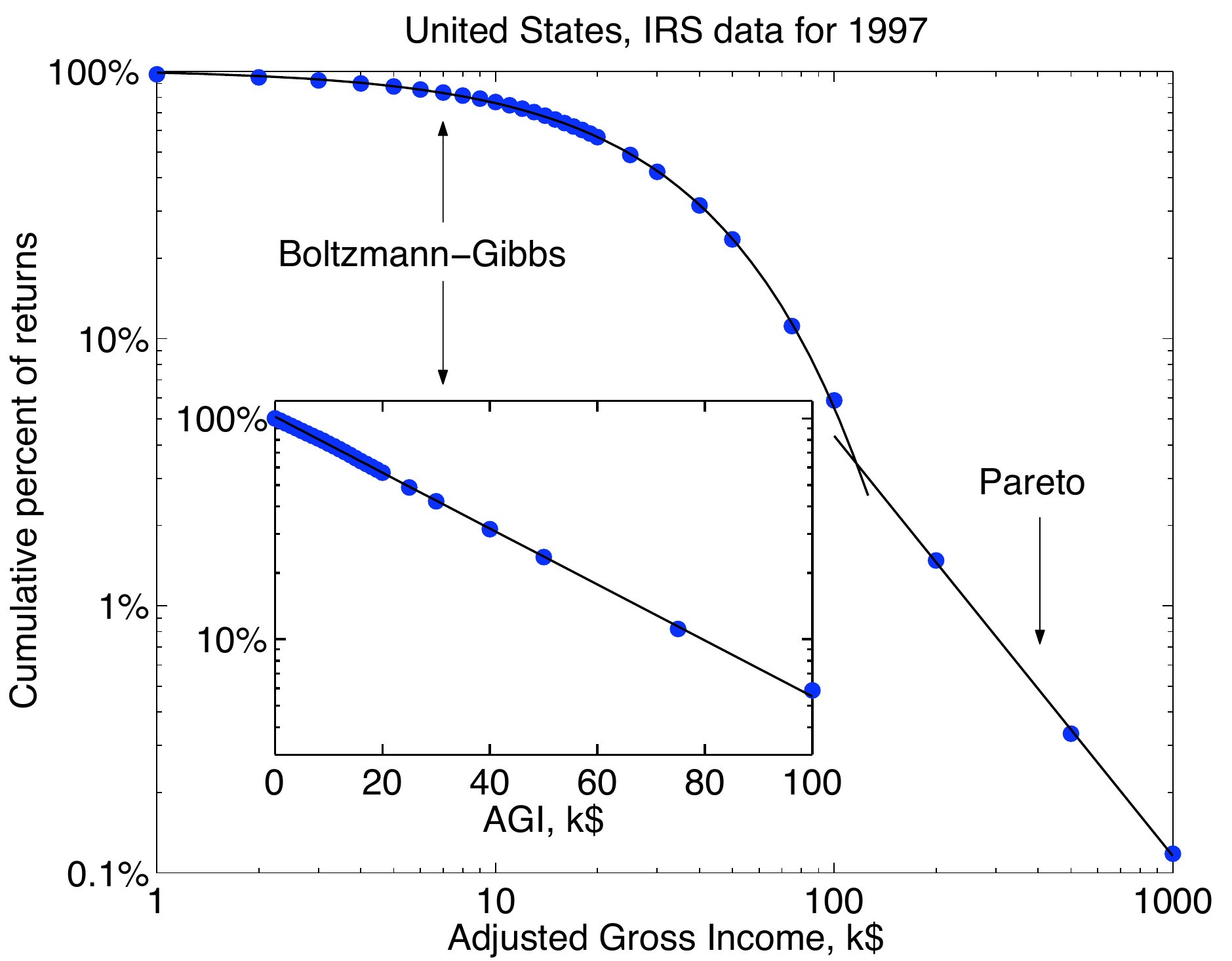}
\caption{Cumulative probability distribution of tax returns for USA in
  1997 shown on log-log (main panel) and log-linear (inset) scales 
  \cite{Dragulescu-2003}.  Points represent the IRS data, and solid lines
  are fits to the exponential and power-law functions.}
\label{Fig:income1997}
\end{figure}

The upper limit of the income data analyzed by \textcite{Dragulescu-2001a} was about 120 k\$/year.  Subsequently, \textcite{Dragulescu-2001b,Dragulescu-2003} analyzed income data up to 1 M\$/year and found that the upper tail of the distribution follows a power law.  Thus, income distribution in the USA has a two-class structure, as shown in Fig.\ \ref{Fig:income1997}.  This figure shows the cumulative distribution function $C(r)=\int_r^\infty P(r')\,dr'$, where $P(r)$ is the probability density, and the variable $r$ denotes income.  When $P(r)$ is an exponential or a power-law function, $C(r)$ is also an exponential or a power-law function.
The straight line on the log-linear scale in the inset of Fig.\ \ref{Fig:income1997} demonstrates the exponential Boltzmann-Gibbs law for the lower class, and the straight line on the log-log scale in the main panel illustrates the Pareto power law for the upper class.  The intersection point between the exponential and power-law fits in Fig.\ \ref{Fig:income1997} defines the boundary between the lower and upper classes.  For 1997, the annual income separating the two classes was about 120~k\$.  About 3\% of the population belonged to the upper class, and 97\% belonged to the lower class.     Although the existence of social classes has been known since Karl Marx, it is interesting that they can be straightforwardly identified by fitting the empirical data with simple mathematical functions.  The exponential distribution (\ref{P(w)}) applies only to the wages of workers in the lower class, whereas the upper tail represents capital gains and other profits of the firms owners.

%%%%%%%%%%%%%%%%%%%%%%%%%%%%%%%%%%%%%%%%%%%%%%%%%%%%%%%%%%%%%%%
%%%%%%%%%%%%%%%%%%%%%%%%%%%%%%%%%%%%%%%%%%%%%%%%%%%%%%%%%%%%%%%

\begin{figure}
\includegraphics[angle=-90,width=0.95\linewidth]{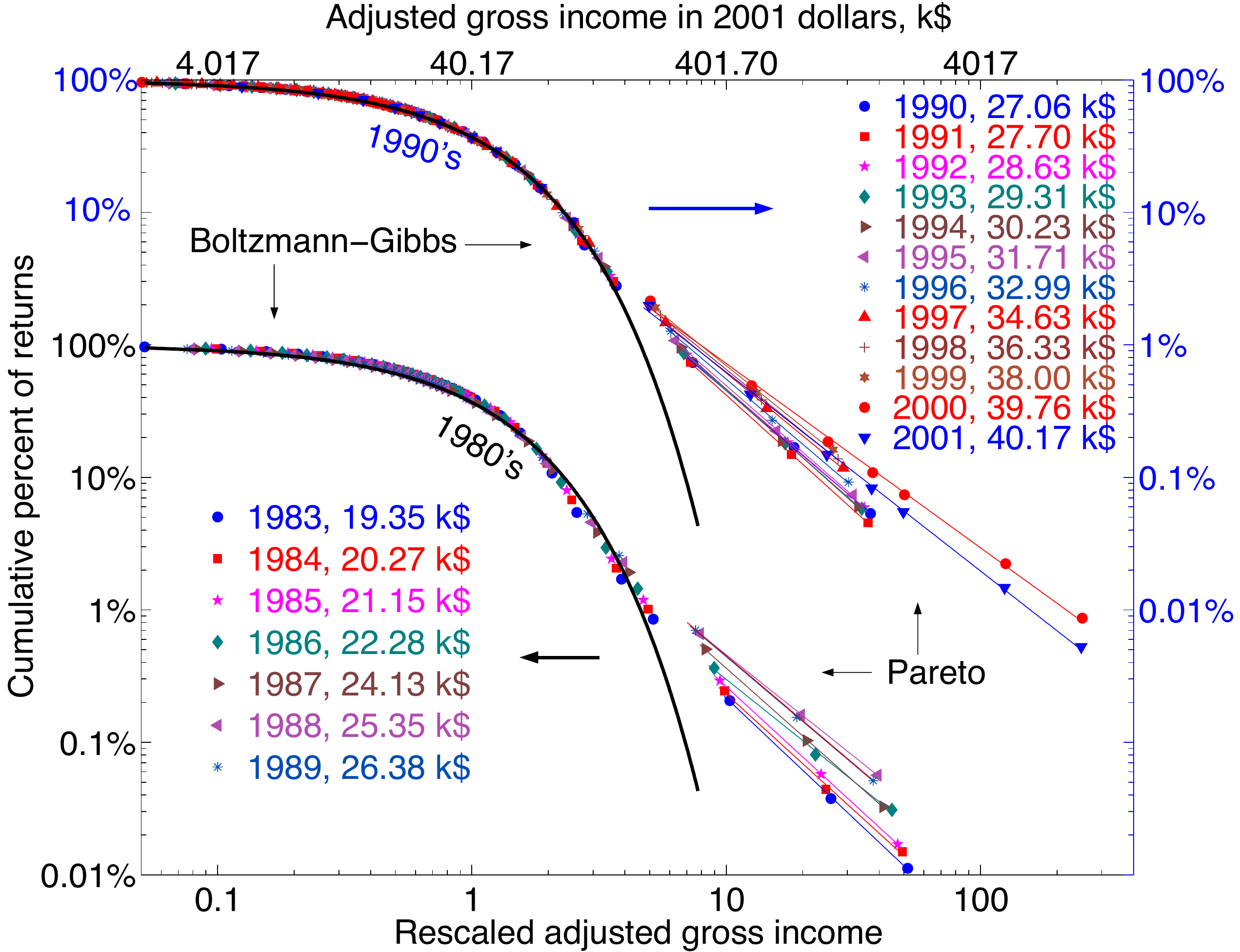}
\caption{Cumulative probability distributions of tax returns for 1983--2001 
  plotted on log-log scale versus $r/T_r$ (the annual income $r$ normalized by
  the average income $T_r$ in the exponential part of the
  distribution) \cite{Silva-2005}.  The columns of numbers give the values of
  $T_r$ for the corresponding years.}
\label{Fig:income}
\end{figure}

\textcite{Silva-2005} studied historical evolution of income distribution in
the USA during 1983--2001 using the IRS data.  The structure of income
distribution was found to be qualitatively similar for all years, as
shown in Fig.\ \ref{Fig:income}.  The average income in nominal
dollars has approximately doubled during this time interval.  So, the
horizontal axis in Fig.\ \ref{Fig:income} shows the normalized income
$r/T_r$, where the income temperature $T_r$ was obtained by
fitting the exponential part of the distribution for each year.
The values of $T_r$ are shown in Fig.\ \ref{Fig:income}.  The plots
for the 1980s and 1990s are shifted vertically for clarity.  We
observe that the data points for the lower class collapse on the same exponential curve for all years.  This demonstrates that the relative income distribution for the
lower class is extremely stable and does not change in time, despite
gradual increase in the average income in nominal dollars.  This
observation confirms that the lower class is in a statistical ``thermal'' equilibrium.

%%%%%%%%%%%%%%%%%%%%%%%%%%%%%%%%%%%%%%%%%%%%%%%%%%%%%%%%%%%%%%%%%%%
%%%%%%%%%%%%%%%%%%%%%%%%%%%%%%%%%%%%%%%%%%%%%%%%%%%%%%%%%%%%%%%%%%%%%%

\begin{figure}
\includegraphics[width=0.9\linewidth]{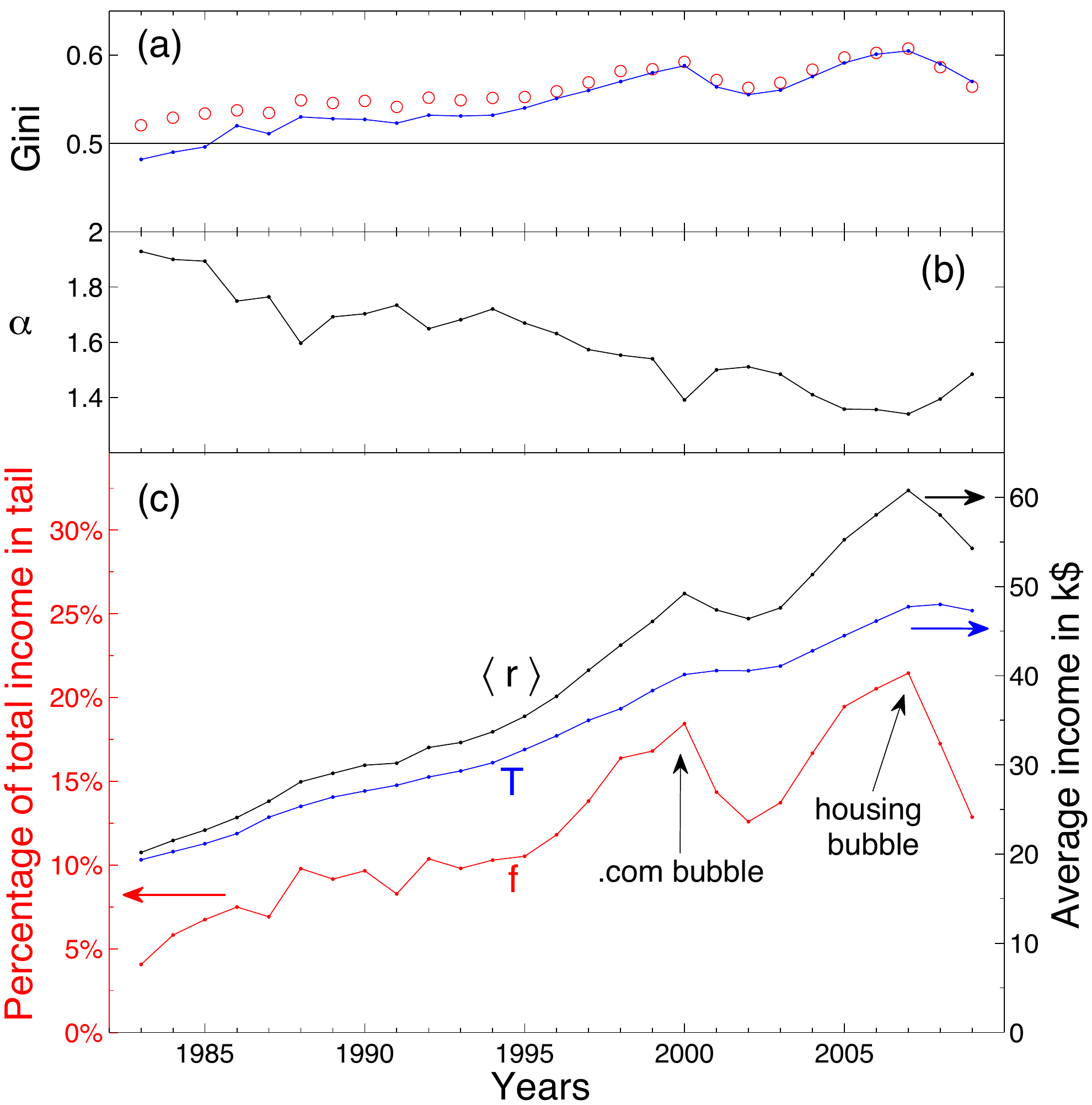}
\caption{(a) The Gini coefficient $G$ for income distribution in the USA in 1983--2009 (connected line), compared with the theoretical formula $G=(1+f)/2$ (open circles).  (b) The exponent $\alpha$ of the power-law tail of income distribution.  (c) The average income $\langle r\rangle$ in the whole system, the average income $T_r$ in the lower class (the temperature of the exponential part), and the fraction of income $f$, Eq.~(\ref{f}), going to the upper tail \cite{Banerjee-2010}.}
\label{Fig:f}
\end{figure}

On the other hand, Fig.\ \ref{Fig:income} also shows that income
distribution of the upper class does not rescale and significantly
changes in time.  \textcite{Banerjee-2010} found that the exponent
$\alpha$ of the power law $C(r)\propto1/r^\alpha$ has decreased from 1.8
in 1983 to 1.3 in 2007, as shown in Panel (b) of Fig.~\ref{Fig:f}.  This means that the upper tail has become ``fatter''.  Another informative parameter is the fraction $f$ of the total income in the system going to the upper class \cite{Silva-2005,Banerjee-2010}:
\begin{equation}
  f=\frac{\langle r\rangle-T_r}{\langle r\rangle}.
\label{f}
\end{equation}
Here $\langle r\rangle$ is the average income of the whole
population, and the temperature $T_r$ is the average income in the
exponential part of the distribution.  Eq.\ (\ref{f}) gives a
well-defined measure of the deviation of the actual income
distribution from the exponential one and, thus, of the fatness of
the upper tail.  Panel (c) in Fig.\ \ref{Fig:f} shows historical evolution of the parameters $\langle r\rangle$, $T_r$, and $f$ \cite{Banerjee-2010}.  We observe that $T_r$ has been increasing approximately linearly in time (this increase mostly represents inflation). In contrast, $\langle r\rangle$ had sharp peaks in 2000 and 2007 coinciding with the heights of speculative bubbles in financial markets.  The fraction $f$ has been increasing for the last 20 years and reached maxima exceeding 20\% in the years 2000 and 2007, followed by sharp drops.  We conclude that speculative bubbles greatly increase overall income inequality by increasing the fraction of income going to the upper class.  When bubbles collapse, income inequality decreases.  Similar results were found for Japan by \textcite{Fujiwara-2003,Aoyama-2003}.

%%%%%%%%%%%%%%%%%%%%%%%%%%%%%%%%%%%%%%%%%%%%%%%%%%%%%%%%%%%%%%%%%%
%%%%%%%%%%%%%%%%%%%%%%%%%%%%%%%%%%%%%%%%%%%%%%%%%%%%%%%%%%%%%%%%%%

\begin{figure}
\includegraphics[angle=-90,width=0.78\linewidth]{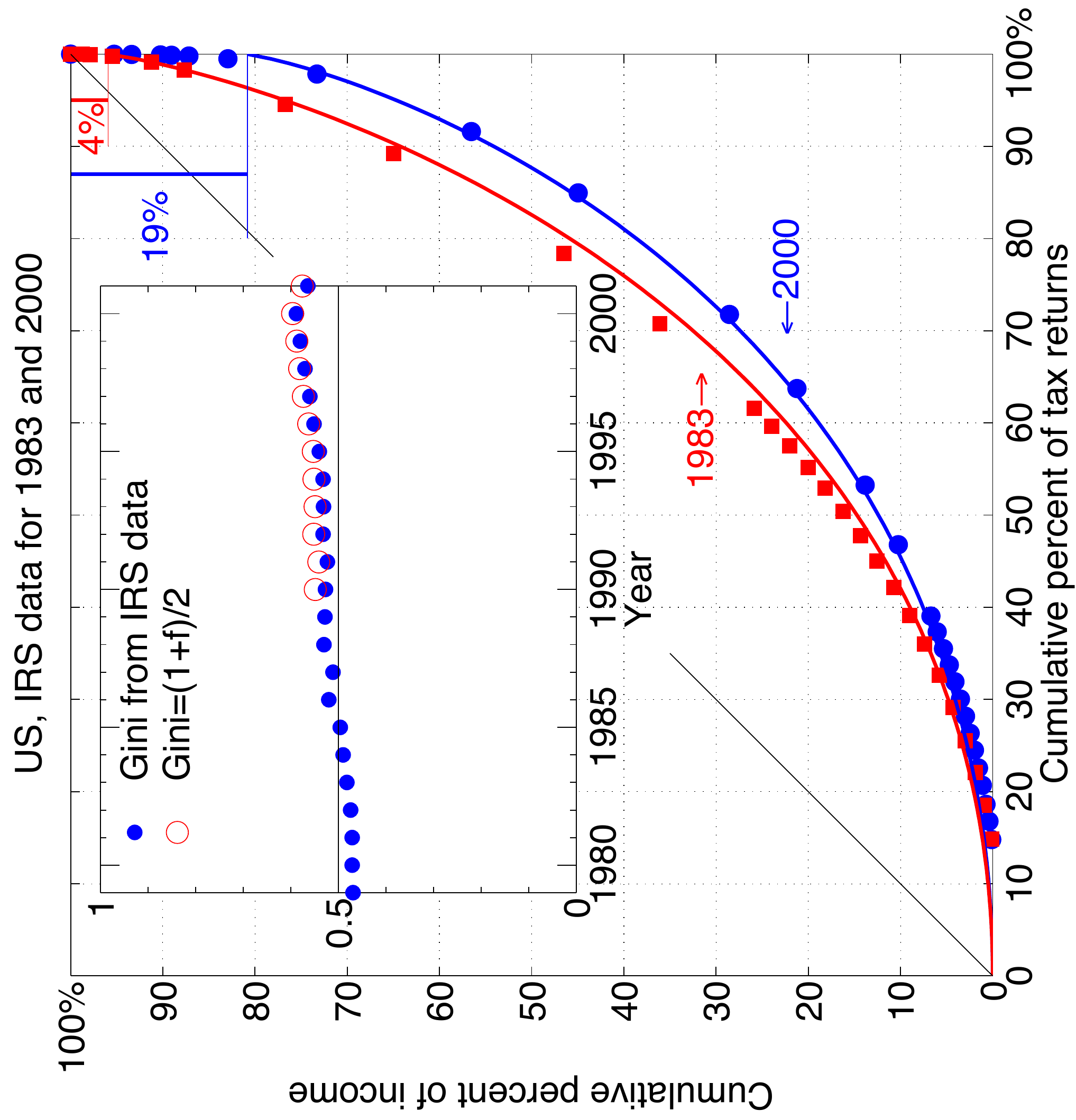}
\caption{\textit{Main panel:} Lorenz plots for income distribution in
  1983 and 2000 \cite{Silva-2005}.  The data points are from the IRS, and 
  the theoretical curves represent Eq.\ (\ref{Lorenz}) with the parameter 
  $f$ deduced from Eq.~(\ref{f}).  \textit{Inset:} The closed circles are the
  IRS data for the Gini coefficient $G$, and the
  open circles show the theoretical formula $G=(1+f)/2$.}
\label{Fig:Lorenz}
\end{figure}

Income inequality can be also characterized by the Lorenz curve and the Gini coefficient \cite{Kakwani-book}.  The Lorenz curve is defined in terms of the two coordinates $x(r)$ and $y(r)$ depending on a parameter $r$:
\begin{equation}
  x(r)=\int_0^r P(r')\,dr',\quad
  y(r)=\frac{\int_0^r r' P(r')\,dr'}{\int_0^\infty r' P(r')\,dr'}.
\label{xy}
\end{equation}
The horizontal coordinate $x(r)$ is the fraction of the population
with income below $r$, and the vertical coordinate $y(r)$ is the
fraction of the total income this population accounts for.  As $r$ changes
from 0 to $\infty$, $x$ and $y$ change from 0 to 1 and parametrically
define a curve in the $(x,y)$ plane.  For for an exponential distribution $P(r)=c\exp(-r/T_r)$, the Lorenz curve is \cite{Dragulescu-2001a}
\begin{equation}
  y=x+(1-x)\ln(1-x).
\label{Lorenz-exp}
\end{equation}
Fig.\ \ref{Fig:Lorenz} shows the Lorenz curves for 1983 and 2000 computed from the IRS data \cite{Silva-2005}.  The data points for 1983 (squares) agree reasonably well with Eq.~(\ref{Lorenz-exp}) for the exponential distribution, which is shown by the upper curve.  In contrast, the fraction $f$ of income in the upper tail becomes so large in 2000 that Eq.~(\ref{Lorenz-exp}) has to be modified as follows \cite{Dragulescu-2003,Silva-2005}
\begin{equation}
  y=(1-f)[x+(1-x)\ln(1-x)]+f\,\Theta(x-1).
\label{Lorenz}
\end{equation}
The last term in Eq.\ (\ref{Lorenz}) represents the vertical jump of
the Lorenz curve at $x=1$, where a small percentage of population
in the upper class accounts for a substantial fraction $f$ of the
total income.  The lower curve in Fig.\ \ref{Fig:Lorenz} shows that
Eq.\ (\ref{Lorenz}) fits the data points for 2000 (circles) very well.

%%%%%%%%%%%%%%%%%%%%%%%%%%%%%%%%%%%%%%%%%%%%%%%%%%%%%%%%%%%%%%%%%%
%%%%%%%%%%%%%%%%%%%%%%%%%%%%%%%%%%%%%%%%%%%%%%%%%%%%%%%%%%%%%%%%%%

The deviation of the Lorenz curve from the straight diagonal line in
Fig.\ \ref{Fig:Lorenz} is a measure of income inequality.  Indeed, if everybody had the same income, the Lorenz curve would be the diagonal line, because the fraction of income would be proportional to the fraction of the population.  The standard measure of income inequality is the Gini coefficient $0\leq G\leq1$, which is defined as the area between the Lorenz curve and the diagonal line, divided by the area of the triangle beneath the diagonal line
\cite{Kakwani-book}.  It was shown by \textcite{Dragulescu-2001a} that the Gini coefficient is $G=1/2$ for a purely exponential distribution.  Historical evolution of the empirical Gini coefficient is shown in the inset of Fig.\ \ref{Fig:Lorenz} and in Panel (a) of Fig.~\ref{Fig:f}.  In the first approximation, the values of $G$ are close to the theoretical value 1/2.  However, if we take into account the upper tail using Eq.\ (\ref{Lorenz}), the formula for the Gini coefficient becomes $G=(1+f)/2$ \cite{Silva-2005}.  The inset in Fig.\ \ref{Fig:Lorenz} and panel (a) in Fig.~\ref{Fig:f} show that this formula gives a very good fit of the IRS data starting from 1995 using the values of $f$ deduced from Eq.~(\ref{f}).  The values $G<1/2$ in the 1980s cannot be captured by this formula, because the Lorenz data points for 1983 lie slightly above the theoretical curve in Fig.\ \ref{Fig:Lorenz}.  We conclude that the increase of income inequality after 1995 originates completely from the growth of the upper-class income, whereas income inequality within the lower class remains constant.

%%%%%%%%%%%%%%%%%%%%%%%%%%%%%%%%%%%%%%%%%%%%%%%%%%%%%%%%%%%%%%%%%%
\begin{figure}
\includegraphics[width=0.9\linewidth]{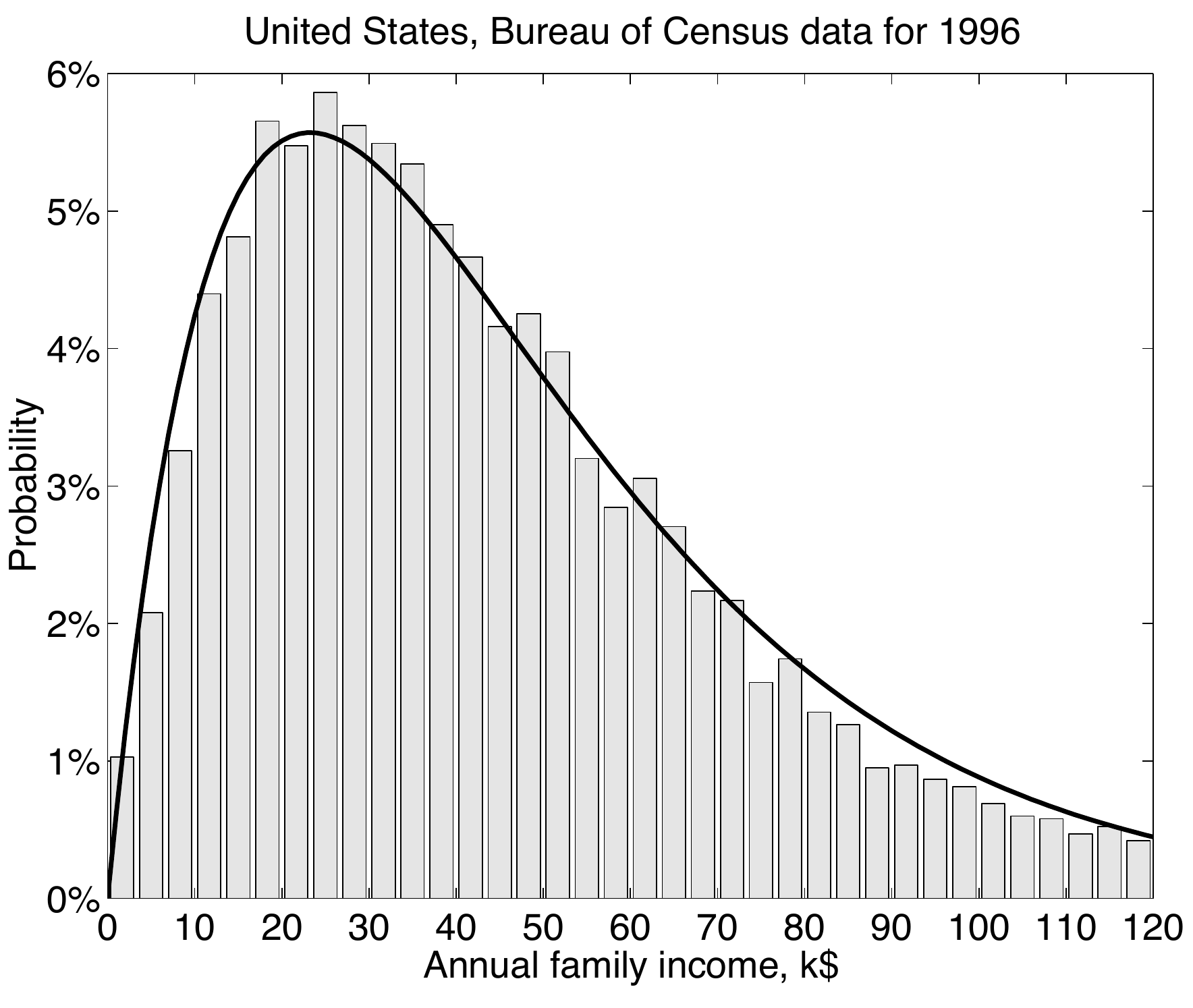}
\caption{\textit{Histogram:} Probability distribution of family income
  for families with two adults, US Census Bureau data 
  \cite{Dragulescu-2001a}.  \textit{Solid line:} Fit to Eq.\ (\ref{family}).}
\label{Fig:income-2}
\end{figure}
%%%%%%%%%%%%%%%%%%%%%%%%%%%%%%%%%%%%%%%%%%%%%%%%%%%%%%%%%%%%%%%%%%

\begin{figure}
\includegraphics[width=0.78\linewidth]{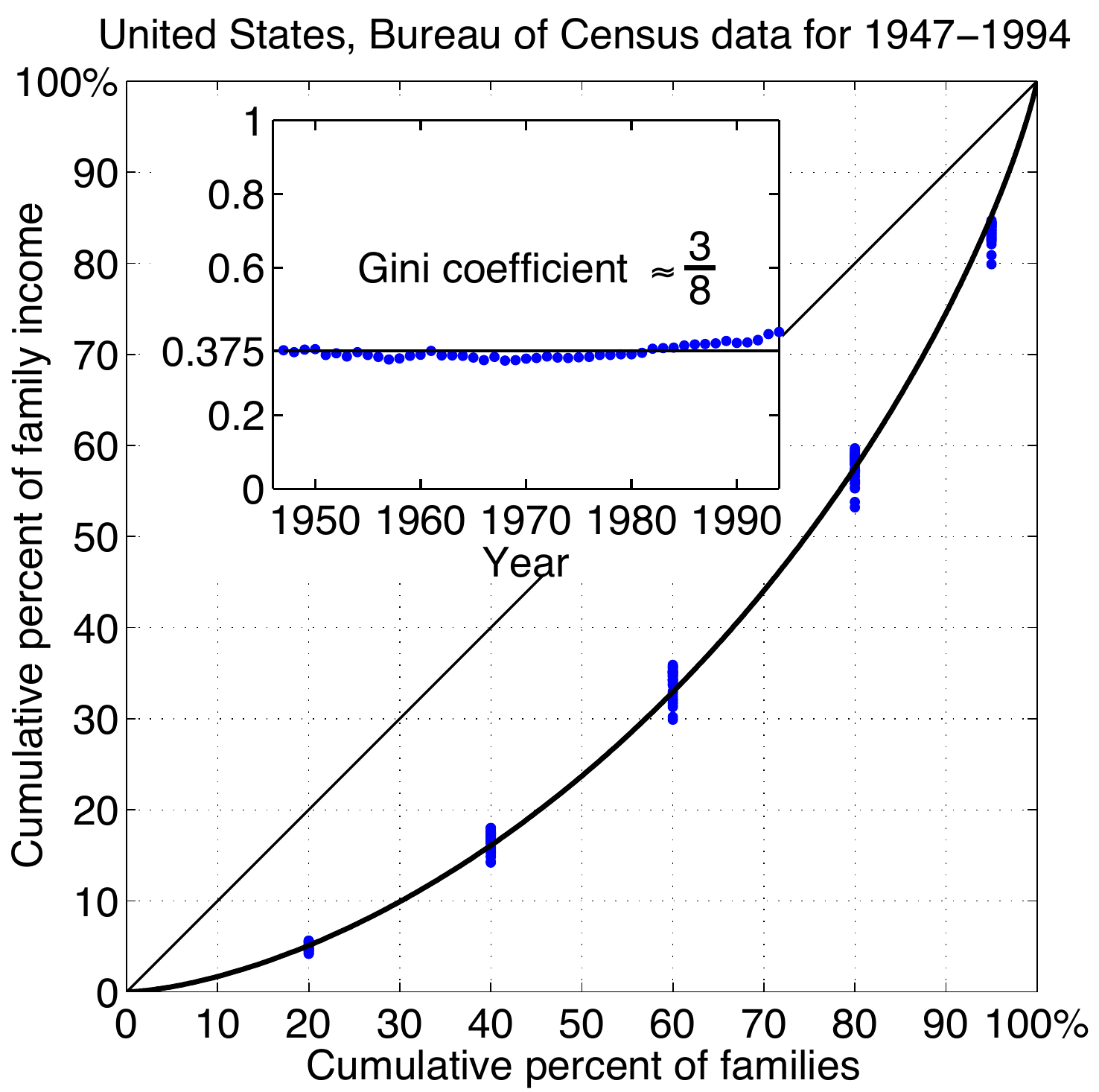}
\caption{\textit{Main panel:} Lorenz plot for family income, calculated
  from Eq.\ (\ref{family}) and compared with the US Census Bureau data 
  points for 1947--1994 \cite{Dragulescu-2001a}.
  \textit{Inset:} Data points from the US Census Bureau for the Gini 
  coefficient for families, compared with the theoretical value
  3/8=37.5\%.}
\label{Fig:Lorenz-2}
\end{figure}

So far, we discussed the distribution of individual income.  An
interesting related question is the distribution $P_2(r)$ of family
income $r=r_1+r_2$, where $r_1$ and $r_2$ are the incomes of spouses.
If individual incomes are distributed exponentially
$P(r)\propto\exp(-r/T_r)$, then
\begin{equation}
  P_2(r)=\int_0^r dr' P(r')P(r-r')=c\,r\exp(-r/T_r),
\label{family}
\end{equation}
where $c$ is a normalization constant.  Fig.\ \ref{Fig:income-2} shows
that Eq.\ (\ref{family}) is in good agreement with the family income
distribution data from the US Census Bureau \cite{Dragulescu-2001a}.
It is assumed in Eq.~(\ref{family}) that incomes of spouses are uncorrelated.  This simple approximation is indeed supported by the scatter plot of incomes of spouses shown in \textcite{Dragulescu-2003}.  The Gini coefficient for the family income distribution (\ref{family}) was analytically calculated by \textcite{Dragulescu-2001a} as $G=3/8=37.5\%$.  Fig.\ \ref{Fig:Lorenz-2} shows the Lorenz quintiles and the Gini coefficient for 1947--1994 plotted from the US Census Bureau data \cite{Dragulescu-2001a}.  The solid line, representing the
Lorenz curve calculated from Eq.\ (\ref{family}), is in good agreement
with the data.  The Gini coefficient, shown in the inset of
Fig.\ \ref{Fig:Lorenz-2}, is close to the calculated value of
$37.5\%$.  Stability of income distribution until the late 1980s was pointed out by \textcite{Levy-1987}, but a very different explanation of this stability was proposed.

%%%%%%%%%%%%%%%%%%%%%%%%%%%%%%%%%%%%%%%%%%%%%%%%%%%%%%%%%%%%%%%%%%
%%%%%%%%%%%%%%%%%%%%%%%%%%%%%%%%%%%%%%%%%%%%%%%%%%%%%%%%%%%%%%%%%%

The exponential distribution of wages (\ref{P(w)}) was derived assuming random realizations of all possible configurations in the absence of specific regulations of the labor market.  As such, it can be taken as an idealized limiting case for comparison with actual distributions in different countries.  If specific measures are implemented for income redistribution, then the distribution may deviate from the exponential one.  An example of such a redistribution was studied by \textcite{Banerjee-2006} for Australia, where $P(r)$ has a sharp peak at a certain income stipulated by government policies.  This is in contrast to the data presented in this section, which indicate that income distribution in the USA is close to the idealized one for a labor market without regulation.  Numerous income distribution studies for other countries are cited in the review paper by \textcite{Yakovenko-2009}.

%%%%%%%%%%%%%%%%%%%%%%%%%%%%%%%%%%%%%%%%%%%%%%%%%%%%%%%%%%%%%%%
\section{Probability distribution of energy consumption}
\label{Sec:energy}
%%%%%%%%%%%%%%%%%%%%%%%%%%%%%%%%%%%%%%%%%%%%%%%%%%%%%%%%%%%%%%%

So far, we discussed how monetary inequality develops for statistical reasons.  Now let us discuss physical aspects of the economy.  Since the beginning of the industrial revolution, rapid technological development of human society has been based on consumption of fossil fuel, such as coal, oil, and gas, accumulated in the Earth for billions of years.  As a result, physical standards of living in modern society are primarily determined by the level of energy consumption per capita.  Now it becomes exceedingly clear that fossil fuel will be exhausted in the not-too-distant future.  Moreover, consumption of fossil fuel releases CO$_2$ into the atmosphere and affects global climate.  These pressing global problems pose great technological and social challenges \cite{Rezai-2012}.

Energy consumption per capita varies widely around the globe.  This heterogeneity is a challenge and a complication for reaching a global consensus on how to deal with the energy problems.  Thus, it is important to understand the origin of the global inequality in energy consumption and characterize it quantitatively.  Here I approach this problem using the method of maximal entropy.

Let us consider an ensemble of economic agents and characterize each agent $j$ by the energy consumption $\epsilon_j$ per unit time.  Notice that here $\epsilon_j$ denotes not energy, but power, which is measured in kiloWatts (kW).  Similarly to Sec.~\ref{Sec:BGphysics}, let us introduce the probability density $P(\epsilon)$, so that $P(\epsilon)\,d\epsilon$ gives the probability to have energy consumption in the interval from $\epsilon$ to $\epsilon+d\epsilon$.  Energy production, based on extraction of fossil fuel from the Earth, is a physically limited resource, which is divided for consumption among the global population.  It would be very improbable to divide this resource equally.  Much more likely, this resource would be divided according to the entropy maximization principle, subject to the global energy production constraint.  Following the same procedure as in Sec.~\ref{Sec:BGphysics}, we arrive to the conclusion that $P(\epsilon)$ should follow the exponential law analogous to Eq.~(\ref{P(e)})
\begin{equation}
  P(\epsilon) = c\, e^{-\epsilon/T_\epsilon}.
\label{P(p)}
\end{equation}
Here $c$ is a normalization constant, and the temperature $T_\epsilon=\langle\epsilon\rangle$ is the average energy consumption per capita.

%%%%%%%%%%%%%%%%%%%%%%%%%%%%%%%%%%%%%%%%%%%%%%%%%%%%%%%%%%%%%%
\begin{figure}
\includegraphics[width=0.9\linewidth]{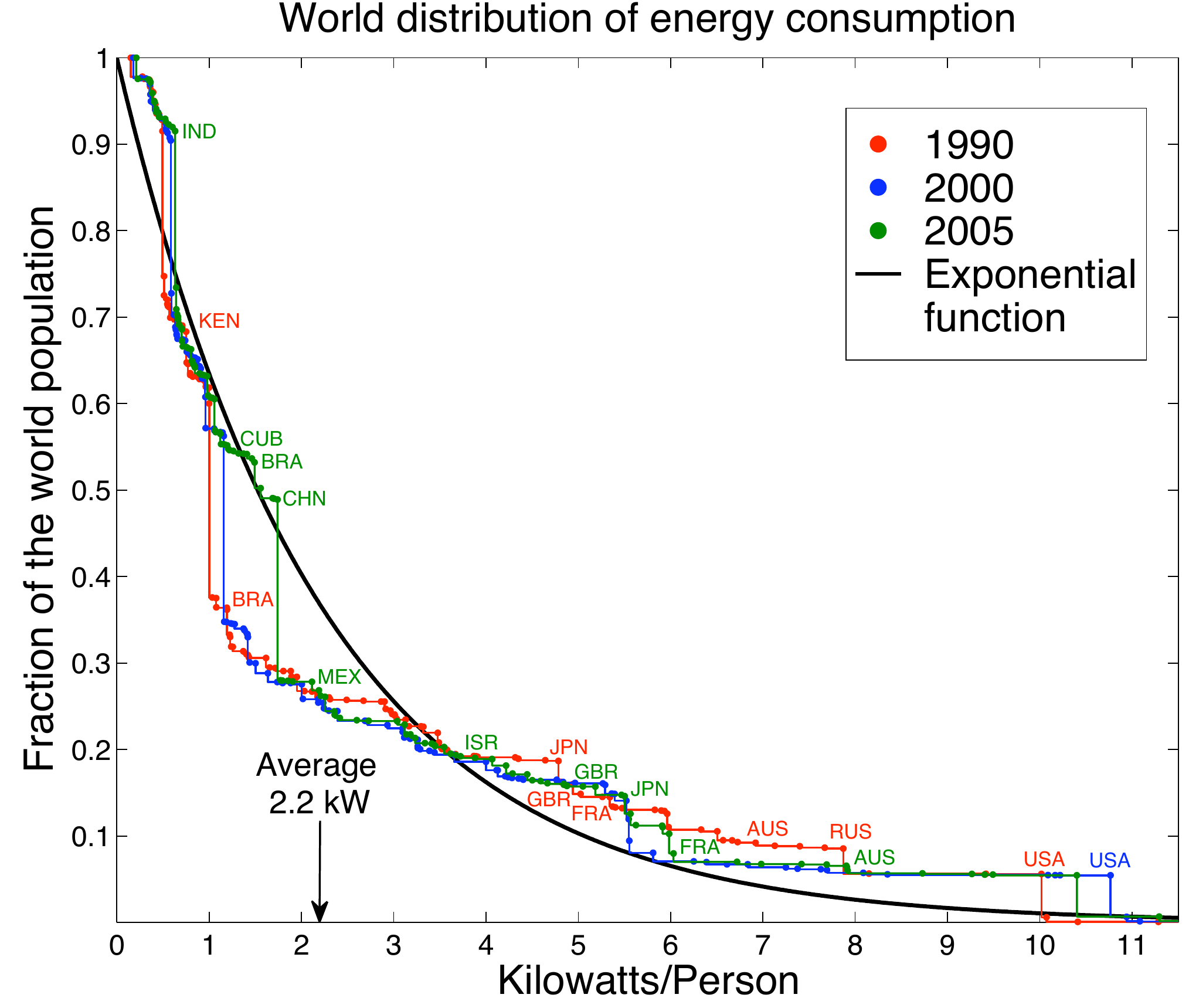}
\caption{Cumulative distribution functions $C(\epsilon)$ for the energy consumption per capita around the world for 1990, 2000, and 2005.  The solid curve is the exponential function (\ref{P(p)}) with the parameter $T_\epsilon=\langle\epsilon\rangle=2.2$ kW.}
\label{fig:eLinlin}
\end{figure}
%%%%%%%%%%%%%%%%%%%%%%%%%%%%%%%%%%%%%%%%%%%%%%%%%%%%%%%%%%%%%%%

Using the data from the World Resources Institute, \textcite{Banerjee-2010} 
constructed the probability distribution of energy consumption per capita around the world and found that it approximately follows the exponential law, as shown in Fig.~\ref{fig:eLinlin}.  The world average energy consumption per capita is $\langle\epsilon\rangle=2.2$ kW, compared with 10 kW in the USA and 0.6 kW in India \cite{Banerjee-2010}.  However, if India and other developing countries were to adopt the same energy consumption level per capita as in the USA, there would be not enough energy resources in the world to do that.  One can argue that the global energy consumption inequality results from the constraint on energy resources, and the global monetary inequality adjusts accordingly to implement this constraint.  Since energy is purchased with money, a fraction of the world population ends up being poor, so that their energy consumption stays limited.

\begin{figure}
\includegraphics[width=0.83\linewidth]{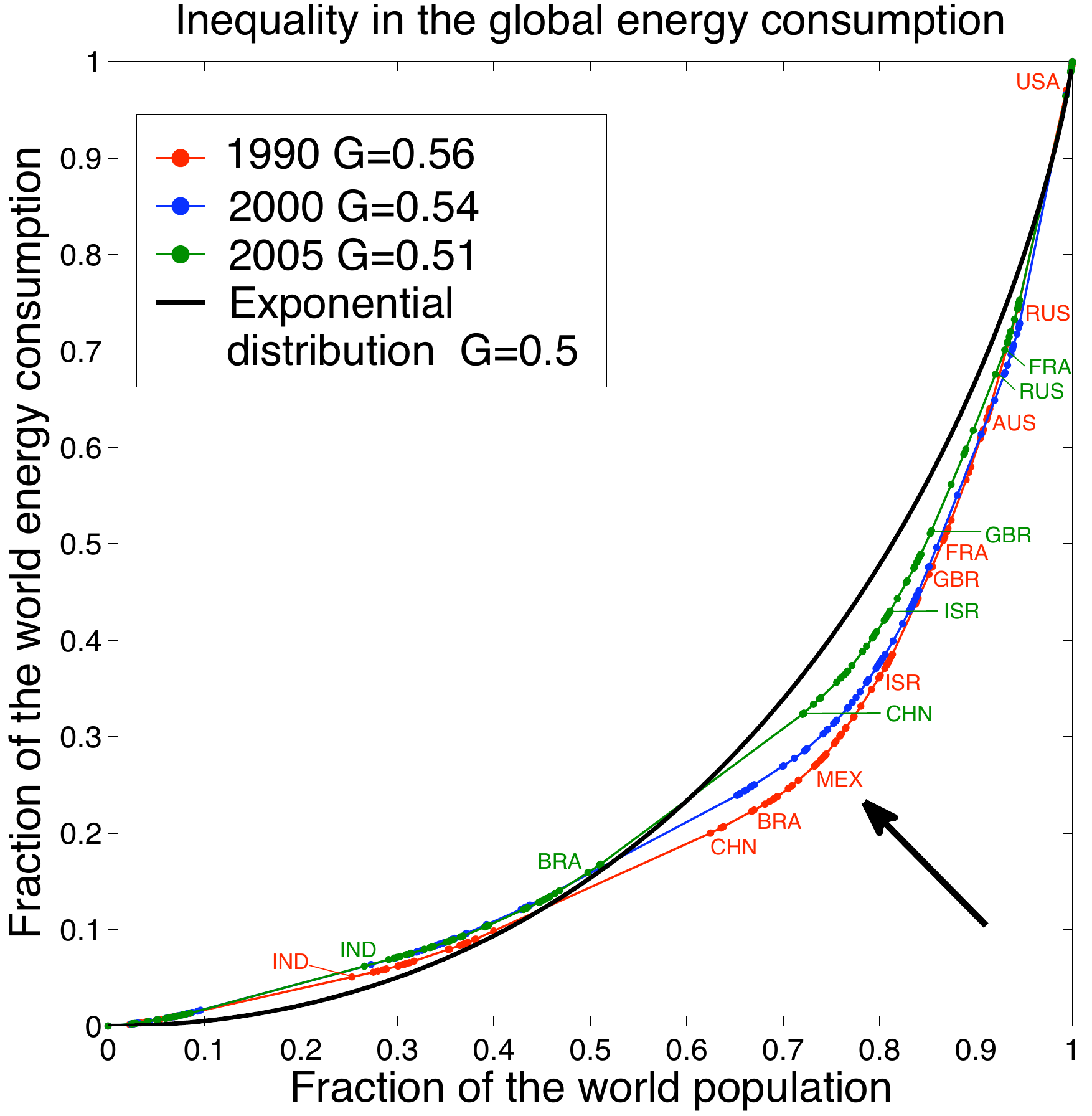}
\caption{Lorenz curves for the energy consumption per capita around the world in 1990, 2000, and 2005, compared with the Lorenz curve  (\ref{Lorenz-exp}) for the exponential distribution.}
\label{fig:elorenz}
\end{figure}

Fig.~\ref{fig:elorenz} shows the Lorenz curves for the global energy consumption per capita in 1990, 2000, and 2005 from \textcite{Banerjee-2010}.  The black solid line is the theoretical Lorenz curve (\ref{Lorenz-exp}) for the exponential distribution (\ref{P(p)}).  In the first approximation, the empirical curves are reasonably close to the theoretical curve, but some deviations are clearly visible.  On the Lorenz curve for 1990, there is a kink or a knee indicated by the arrow, where the slope of the curve changes appreciably.  This point separates developed and developing countries.  Mexico, Brazil, China, and India are below this point, whereas Britain, France, Japan, Australia, Russia, and USA are above.  The slope change  of the Lorenz curve represents a gap in the energy consumption per capita between these two groups of countries.  Thus, the physical difference between developed and developing countries lies in the degree of energy consumption and utilization, rather than in more ephemeral monetary measures, such as dollar income per capita.  However, the Lorenz curve for 2005 is closer to the exponential curve, and the kink is less pronounced.  It means that the energy consumption inequality and the gap between developed and developing countries have decreased, as also confirmed by the decrease in the Gini coefficient $G$ listed in Fig.~\ref{fig:elorenz}.  This result can be attributed to the rapid globalization and stronger mixing of the world economy in the last 20 years.  However, the energy consumption distribution in a well-mixed globalized world economy approaches to the exponential one, rather than to an equal distribution.

%%%%%%%%%%%%%%%%%%%%%%%%%%%%%%%%%%%%%%%%%%%%%%%%%%%%%%%%%%%%%%
%%%%%%%%%%%%%%%%%%%%%%%%%%%%%%%%%%%%%%%%%%%%%%%%%%%%%%%%%%%%%%%

The inherent inequality of the global energy consumption makes it difficult for the countries at the opposite ends of the distribution to agree on consistent measures to address the energy and climate challenges.  While not offering any immediate solutions, I would like to point out that renewable energy has different characteristics than fossil fuel.  Because solar and wind energy is typically generated and consumed locally and not transported on the global scale, it is not subject to the entropy-maximizing redistribution.  Thus, a transition from fossil fuel to renewable energy gives a hope for achieving a more equal global society.

%%%%%%%%%%%%%%%%%%%%%%%%%%%%%%%%%%%%%%%%%%%%%%%%%%%%%%%%%%%%%%%
\section{Conclusions}
\label{Sec:conclusions}
%%%%%%%%%%%%%%%%%%%%%%%%%%%%%%%%%%%%%%%%%%%%%%%%%%%%%%%%%%%%%%%

Statistical approach and entropy maximization are very general and powerful methods for studying big ensembles of various nature, be it physical, biological, economical, or social.  Duncan Foley is one of the pioneers of applying these methods in economics \cite{Foley-1994,Foley-1996,Foley-1999}. 
The exponential distribution of wages, predicted by the statistical equilibrium theory of a labor market \cite{Foley-1996}, is supported by empirical data on income distribution in the USA for the majority of population.  In contrast, the upper tail of income distribution follows a power law and expands dramatically during financial bubbles, which results in a significant increase of the overall income inequality.  The two-class structure of the American society is apparent in the plots of income distribution, where the lower and upper classes are described by the exponential and power-law distributions.  The entropy maximization method also demonstrates how a highly unequal exponential probability distribution of money among the initially equal agents is generated as a results of stochastic monetary transactions between the agents.  These ideas also apply to the global inequality of energy consumption per capita around the world.  The empirical data show convergence to the predicted exponential distribution in the process of globalization.  Global monetary inequality may be a consequence of the constraint on global energy resources, because it limits energy consumption per capita for a large fraction of the world population.

%%%%%%%%%%%%%%%%%%%%%%%%%%%%%%%%%%%%%%%%%%%%%%%%%%%%%%%%%%%%%%%

%%%%%%%%%%%%%%%%%%%%%%%%%%%%%%%%%%%%%%%%%%%%%%%%%%%%%%%%%%%%%%%%
\end{document}